\documentclass[twocolumn,showpacs,superscriptaddress,preprintnumbers,amsmath,amssymb]{revtex4}

\usepackage{xspace}
\usepackage{graphicx}
\usepackage{dcolumn}
\usepackage{bm}

\newcommand{\sk}      {Super-Kamiokande\xspace}

\begin{document}

\preprint{APS/123-QED}

\title{Three flavor neutrino oscillation analysis of atmospheric
neutrinos in Super-Kamiokande \\
}

\newcommand{\icrr}{\affiliation{Kamioka Observatory, Institute for Cosmic Ray Research, University of Tokyo, Kamioka, Gifu, 506-1205, Japan}}
\newcommand{\ncen}{\affiliation{Research Center for Cosmic Neutrinos, Institute for Cosmic Ray Research, University of Tokyo, Kashiwa, Chiba 277-8582, Japan}}
\newcommand{\bu}{\affiliation{Department of Physics, Boston University, Boston, MA 02215, USA}}
\newcommand{\bnl}{\affiliation{Physics Department, Brookhaven National Laboratory, Upton, NY 11973, USA}}
\newcommand{\uci}{\affiliation{Department of Physics and Astronomy, University of California, Irvine, Irvine, CA 92697-4575, USA}}
\newcommand{\csu}{\affiliation{Department of Physics, California State University, Dominguez Hills, Carson, CA 90747, USA}}
\newcommand{\cnu}{\affiliation{Department of Physics, Chonnam National University, Kwangju 500-757, Korea}}
\newcommand{\duke}{\affiliation{Department of Physics, Duke University, Durham, NC 27708 USA}}
\newcommand{\gmu}{\affiliation{Department of Physics, George Mason University, Fairfax, VA 22030, USA}}
\newcommand{\gifu}{\affiliation{Department of Physics, Gifu University, Gifu, Gifu 501-1193, Japan}}
\newcommand{\uh}{\affiliation{Department of Physics and Astronomy, University of Hawaii, Honolulu, HI 96822, USA}}
\newcommand{\ui}{\affiliation{Department of Physics, Indiana University, Bloomington,  IN 47405-7105, USA} }
\newcommand{\kek}{\affiliation{High Energy Accelerator Research Organization (KEK), Tsukuba, Ibaraki 305-0801, Japan}}
\newcommand{\kobe}{\affiliation{Department of Physics, Kobe University, Kobe, Hyogo 657-8501, Japan}}
\newcommand{\kyoto}{\affiliation{Department of Physics, Kyoto University, Kyoto 606-8502, Japan}}
\newcommand{\lanl}{\affiliation{Physics Division, P-23, Los Alamos National Laboratory, Los Alamos, NM 87544, USA}}
\newcommand{\lsu}{\affiliation{Department of Physics and Astronomy, Louisiana State University, Baton Rouge, LA 70803, USA}}
\newcommand{\umd}{\affiliation{Department of Physics, University of Maryland, College Park, MD 20742, USA}}
\newcommand{\MIT}{\affiliation{Department of Physics, Massachusetts Institute of Technology, Cambridge, MA 02139, USA}}
\newcommand{\duluth}{\affiliation{Department of Physics, University of Minnesota, Duluth, MN 55812-2496, USA}}
\newcommand{\miyagi}{\affiliation{Department of Physics, Miyagi University of Education, Sendai,Miyagi 980-0845, Japan}}
\newcommand{\suny}{\affiliation{Department of Physics and Astronomy, State University of New York, Stony Brook, NY 11794-3800, USA}}
\newcommand{\nagoya}{\affiliation{Department of Physics, Nagoya University, Nagoya, Aichi 464-8602, Japan}}
\newcommand{\nagoyaste}{\affiliation{Solar-Terrestrial Environment Laboratory, Nagoya University, Nagoya, Aichi 464-8601, Japan}}
\newcommand{\niigata}{\affiliation{Department of Physics, Niigata University, Niigata, Niigata 950-2181, Japan}}
\newcommand{\osaka}{\affiliation{Department of Physics, Osaka University, Toyonaka, Osaka 560-0043, Japan}}
\newcommand{\okayama}{\affiliation{Department of Physics, Okayama University, Okayama, Okayama 700-8530, Japan}}
\newcommand{\seoul}{\affiliation{Department of Physics, Seoul National University, Seoul 151-742, Korea}}
\newcommand{\shizuokaseika}{\affiliation{International and Cultural Studies, Shizuoka Seika College, Yaizu, Shizuoka 425-8611, Japan}}
\newcommand{\shizuokafukushi}{\affiliation{Department of Informatics in
Social Welfare, Shizuoka University of Welfare, Yaizu, Shizuoka 425-8611, Japan}}
\newcommand{\shizuoka}{\affiliation{Department of Systems Engineering, Shizuoka University, Hamamatsu, Shizuoka 432-8561, Japan}}
\newcommand{\skku}{\affiliation{Department of Physics, Sungkyunkwan University, Suwon 440-746, Korea}}
\newcommand{\tohoku}{\affiliation{Research Center for Neutrino Science, Tohoku University, Sendai, Miyagi 980-8578, Japan}}
\newcommand{\tokyo}{\affiliation{University of Tokyo, Tokyo 113-0033, Japan}}
\newcommand{\tokai}{\affiliation{Department of Physics, Tokai University, Hiratsuka, Kanagawa 259-1292, Japan}}
\newcommand{\tit}{\affiliation{Department of Physics, Tokyo Institute for Technology, Meguro, Tokyo 152-8551, Japan}}
\newcommand{\warsaw}{\affiliation{Institute of Experimental Physics, Warsaw University, 00-681 Warsaw, Poland}}
\newcommand{\uw}{\affiliation{Department of Physics, University of Washington, Seattle, WA 98195-1560, USA}}
\newcommand{\tsukubanow}{\altaffiliation{ Present address: Department of Physics, Univ. of Tsukuba, Tsukuba, Ibaraki 305 8577, Japan}}
\newcommand{\okayamanow}{\altaffiliation{ Present address: Department of Physics, Okayama University, Okayama 700-8530, Japan}}
\newcommand{\marylandnow}{\altaffiliation{ Present address: University of Maryland School of Medicine, Baltimore, MD 21201, USA}}
\newcommand{\triunfnow}{\altaffiliation{ Present address: TRIUMF, Vancouver, British Columbia V6T 2A3, Canada}}
\newcommand{\icrrnow}{\altaffiliation{ Present address: Kamioka Observatory, Institute for Cosmic Ray Research, University of Tokyo, Kamioka, Gifu, 506-1205, Japan}}
\newcommand{\pennnow}{\altaffiliation{ Present address: Center for Gravitational Wave Physics, Pennsylvania State University, University Park, PA 16802, USA}}
%
\author{J.Hosaka}\icrr
\author{K.Ishihara}\icrr
\author{J.Kameda}\icrr
\author{Y.Koshio}\icrr
\author{A.Minamino}\icrr
\author{C.Mitsuda}\icrr
\author{M.Miura}\icrr
\author{S.Moriyama}\icrr
\author{M.Nakahata}\icrr
\author{T.Namba}\icrr
% ...
\collaboration{The Super-Kamiokande Collaboration}\noaffiliation

\date{\today}

\begin{abstract}

We report on the results of a three-flavor oscillation analysis
using Super-Kamiokande~I atmospheric neutrino data,
with the assumption of one mass scale dominance ($\Delta m_{12}^2$$=$0).
No significant 
flux change
due to matter effect, which occurs
when neutrinos propagate inside the Earth for
$\theta_{13}$$\neq$0,
has been seen either in a multi-GeV $\nu_e$-rich sample or in a 
$\nu_\mu$-rich sample.
Both normal and inverted mass hierarchy hypotheses are tested and
both are consistent with observation.
Using Super-Kamiokande data only, 2-dimensional 90 \% confidence allowed
regions are obtained: mixing angles are constrained to
$\sin^2\theta_{13} < 0.14$ and $0.37 < \sin^2\theta_{23} < 0.65$ for the 
normal mass hierarchy.
Weaker constraints, 
$\sin^2\theta_{13} < 0.27$ and $0.37 < \sin^2\theta_{23} < 0.69$,
are obtained for the inverted mass hierarchy case.

\end{abstract}

\maketitle

\section{\label{sec:intro}Introduction} 

\def\ket#1{|#1\rangle}


Recently a number of experiments have shown evidence for oscillations
of atmospheric~\cite{Becker-Szendy:1992hq,
Fukuda:1994mc, Ambrosio:2003yz, Allison:2005dt, Ashie:2005ik, Adamson:2005qc},
solar~\cite{Smy:2003jf, Ahmed:2003kj},
reactor~\cite{Apollonio:2002gd},
and accelerator neutrinos~\cite{Aliu:2004sq}.


In the standard oscillation picture, the three neutrino flavor eigenstates
are related to the mass eigenstates by a $3\times 3$ unitary mixing matrix $U$: 
\begin{equation}
\ket{\nu_\alpha} = \sum_{i=1}^{3} U_{\alpha i} \ket{\nu_i}.
\label{eq:nuosc}
\end{equation}
In this picture, 
neutrino oscillations can be described by six
parameters: two independent 
$\Delta m^2$$_{ij}$ 
($\Delta m^2_{12}$, $\Delta m^2_{23}$), 
three mixing angles ($\theta_{12}$, $\theta_{23}$,
$\theta_{13}$), and a CP-violating phase $\delta$.  The mixing matrix
$U$ of Eq. \ref{eq:nuosc} can be written as a product of three
rotations, each described by one of the mixing angles:



\begin{eqnarray}
\rm{U} & = & 
\left(
\begin{array}{ccc}
1 & 0 & 0\\ 
0 & c_{23} & s_{23} \\  
0 & -s_{23} & c_{23} 
\end{array} \right)
\left(
\begin{array}{ccc}
c_{13} & 0 & s_{13}e^{-i\delta}\\ 
0 & 1 & 0 \\  
-s_{13}e^{i\delta} & 0 & c_{13} 
\end{array} \right) \nonumber \\
&& \times 
\left(
\begin{array}{ccc}
c_{12} & s_{12} & 0\\ 
-s_{12} & c_{12} & 0 \\  
0 & 0 & 1 
\end{array} \right)
\end{eqnarray}

\noindent
where ``$s$'' represents sine of the mixing angle and ``$c$'' represents
cosine.

The ``1-2'' matrix describes solar mixing; the ``2-3'' matrix
describes atmospheric neutrino mixing.  The ``1-3'' mixing is known to
be small; the best current limits on $\theta_{13}$ come from the CHOOZ
experiment~\cite{Apollonio:1999ae}.

As yet, the Super-Kamiokande atmospheric neutrino oscillation fits
have been done within a two-flavor oscillation
framework~\cite{Ashie:2004mr, Ashie:2005ik}.  In this paper, we
explore the atmospheric data in the context of a three-flavor analysis.


Among the remaining problems in neutrino physics that can be 
answered by oscillation experiments are whether $\theta_{13}$ 
is non-zero, and whether the hierarchy is normal or inverted, $i.e.$
whether $\Delta m^2_{23}$ is positive or negative.
At baselines and energies appropriate for atmospheric neutrinos, the
signature of a non-zero $\theta_{13}$ is 
a matter-enhanced excess of upward-going electron-like
events
and possible additional small rate changes of upward-going muon-like events
with respect to two-flavor $\nu_\mu \rightarrow \nu_\tau$
transition.
The expected effects on electron-like (and muon-like) event rates 
differ for normal and inverted mass hierarchy cases because 
the matter effect
and the cross section differ for \(\nu_e\) and \(\bar{\nu}_e\).


\section{\label{sec:oscillation}Three neutrino oscillation with one mass scale dominant} 

%
%
%
%
%
%

In general, neutrino oscillations are driven by differences of
squared masses, \(m_1^2, m_2^2, m_3^2\).  We have adopted the so-called
``one mass scale dominance" framework:

\begin{equation}
|m_2^2-m_1^2| << |m_3^2-m_{1,2}^2| 
\label{eqn:sqmassdiff}
\end{equation}
This approximation is supported by experimental observations
of 
solar, reactor, atmospheric and accelerator neutrino oscillations.
The advantage of this framework is that the number of parameters involved
in neutrino oscillations is reduced to three:
two mixing angles (\(\theta_{23}, \theta_{13}\)) and one 
mass squared difference \(\Delta m^2\),
\begin{equation}
\Delta m^2 \equiv m_3^2-m_{1,2}^2. 
\label{eqn:sqmassdiff2}
\end{equation}
The ignored oscillation effects driven by the smaller mass difference
$\Delta m^2_{12} \equiv |m_2^2-m_1^2|$, 
which might be observed for a neutrino energy of
a few hundred MeV, are known to be greatly reduced in the case that
the initial flavor flux ratio of $\nu_\mu/\nu_e$ is 2
and $\sin^22\theta_{23}$ is close to one \cite{Peres:2003wd}.
For oscillations driven by the dominant $\Delta m^2$,
a similar ``screening'' effect holds, but more weekly,
because the flavor ratio starts deviating from 2 at 1 GeV
and reaches $\sim$ 3 at 10 GeV.
Equation \ref{eqn:sqmassdiff} can hold both  for
\(m_{1,2}^2 << m_3^2 \) (normal mass hierarchy) and
\(m_3^2 << m_{1,2}^2\) (inverted mass hierarchy).  
We present tests of both cases in this paper.

In this framework, 
in the case of non-zero \(\theta_{13}\) the neutrino transition
and survival
probabilities in vacuum,
valid for down-going atmospheric neutrinos, are expressed as 
\begin{eqnarray}
\mbox{P}(\nu_e \rightarrow \nu_e) &=& 1- \sin^2 2\theta_{13}
      \sin^2 \left(\frac{1.27 \Delta m^2 L}{E}\right) \nonumber \\
\mbox{P}(\nu_\mu \rightarrow \nu_e) &=& \mbox{P}(\nu_e \rightarrow \nu_\mu) \nonumber \\
&=& \sin^2 \theta_{23} \sin^2 2 \theta_{13}
      \sin^2 \left(\frac{1.27 \Delta m^2 L}{E}\right) \nonumber \\
\mbox{P}(\nu_\mu \rightarrow \nu_\mu) &=& 1 \nonumber \\
&& - 4 \cos^2 \theta_{13}
\sin^2 \theta_{23} ( 1-\cos^2 \theta_{13} \sin^2 \theta_{23}) \nonumber \\
&&\times 
\sin^2 \left(\frac{1.27 \Delta m^2 L}{E}\right)
\label{eqn:oscillation-vacuum}
\end{eqnarray}
where \(L\) is neutrino travel length in km from the neutrino
production point in the atmosphere,
\(E\) is neutrino energy in GeV, and $\Delta m^2$ in eV$^2$.
In the limit of zero \(\theta_{13}\), 
these equations reduce to pure 
\(\nu_\mu \leftrightarrow \nu_\tau\) two-flavor oscillation.
Because \(\mbox{P}(\nu_e \rightarrow \nu_e)\) 
in Eq. \ref{eqn:oscillation-vacuum} 
is a function of \(\sin^22\theta_{13}\),
both \(\sin^2\theta_{13} \sim 0\) and \(\sin^2\theta_{13} \sim 1\) can
 satisfy 
electron flavor disappearance constraints
from reactor neutrino experiments.
However, \(\sin^2\theta_{13} \sim 1\) is inconsistent with 
an observed large deficit of atmospheric \(\nu_\mu\) 
because \(\mbox{P}(\nu_\mu \rightarrow \nu_\mu)\) 
in Eq. \ref{eqn:oscillation-vacuum} becomes \(\sim 1\)
for this case.
Therefore, \(\sin^2\theta_{13}>0.5\) is not discussed here.

For neutrinos traversing the Earth,
oscillation probability is calculated taking into account 
Earth's matter potential due to the forward scattering amplitude
of charged current $\nu_e$ and $\bar{\nu}_e$ interactions
\cite{MS,MS2,Wolfenstein:1979ni}.
We adopted a model in which the Earth is well-approximated by
four layers of a constant matter density
(core~1: R$\leq$1221~km, $\rho$$=$13.0~g/cm$^3$,
core~2: 1221$<$R$\leq$3480~km, $\rho$$=$11.3~g/cm$^3$,
mantle: 3480$<$R$\leq$5701~km, $\rho$$=$5.0~g/cm$^3$,
surface: 5701$<$R$\leq$6371~km, $\rho$$=$3.3~g/cm$^3$).
The method of calculating matter oscillation probabilities 
in constant density is based on \cite{Barger:1980tf}.

In constant matter density, 
\(\mbox{P}(\nu_\mu \rightarrow \nu_e)\)
can be described by replacing the mixing angle \(\theta_{13}\)
and \(\Delta m^2\) in Eq. \ref{eqn:oscillation-vacuum} 
as
\begin{eqnarray}
\mbox{P}(\nu_\mu \rightarrow \nu_e) &=& \mbox{P}(\nu_e \rightarrow \nu_\mu) \nonumber \\
&=& \sin^2 \theta_{23} \sin^2 2 \theta_{13}^M
      \sin^2 \left(\frac{1.27 \Delta m^2_{M} L}{E}\right) \nonumber \\
\label{eqn:oscillation-matter}
\end{eqnarray}
and the effective mixing angle is
\begin{eqnarray}
\sin^2 2\theta_{13}^M &=& 
      \frac{\sin^2 2\theta_{13}}{(\cos2\theta_{13}-A_{CC}/\Delta m^2)^2 
      + \sin^22\theta_{13}} \nonumber \\
A_{CC} &=& 2\sqrt{2}G_FN_ep \nonumber \\
\label{eqn:effective-mixing}
\end{eqnarray}
where \(G_F\) is the Fermi constant, \(N_e\) is the electron densities
in the medium and \(p\) is neutrino momentum \cite{Giunti:1997fx}.
The matter potential term \(A_{CC}\) has the same absolute value, but opposite
sign for neutrinos and anti-neutrinos.
Due to the matter effect, 
the Mikheyev-Smirnov-Wolfenstein (MSW) resonance happens in that 
the \(\nu_\mu \rightarrow \nu_e\) oscillation
probability in Eq. \ref{eqn:oscillation-matter} and \ref{eqn:effective-mixing}
becomes large at \(3\sim10\) GeV neutrino energy
even for the case of small \(\theta_{13}\)
\cite{Akhmedov:1998xq,Chizhov:1998ug,Gonzalez-Garcia:2002mu,Bernabeu:2003yp}.
To demonstrate the behavior of \(\nu_e\) oscillations,
\begin{figure}
\begin{center}
\includegraphics[width=60mm]{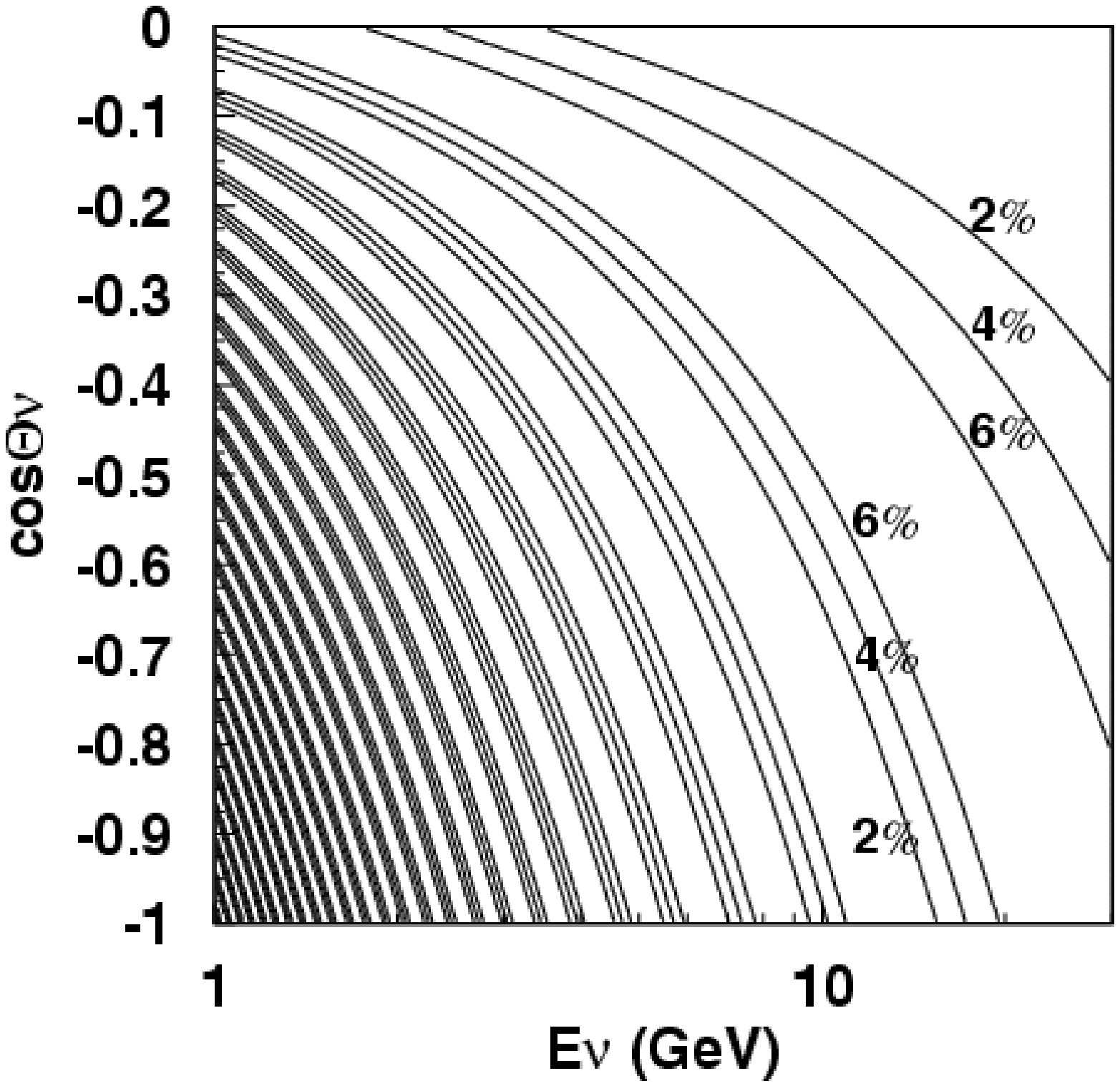}
\includegraphics[width=60mm]{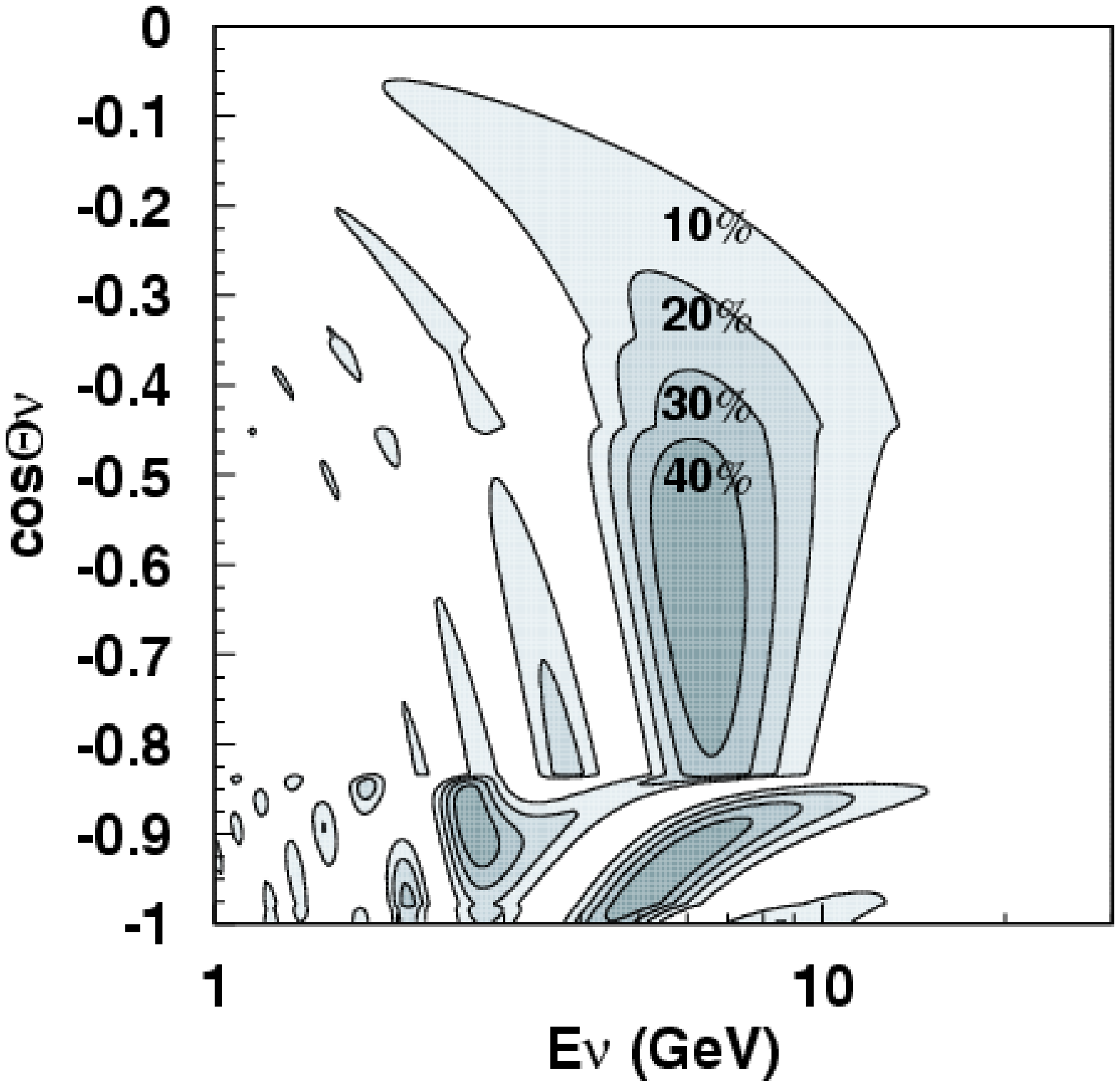}
\caption{
Oscillation probability of $\nu_e \rightarrow \nu_\mu$ 
(or $\nu_\mu \rightarrow \nu_e$) transition.
For both figures, the horizontal axis shows neutrino energy and the
vertical axis shows the zenith angle of neutrino direction;
\(\cos\Theta_\nu=-1\) and \(\cos\Theta_\nu=0\) correspond to
vertically upward and horizontal directions, respectively.
Angles with \(\cos\Theta_\nu<-0.84\) correspond to neutrinos passing through
the earth core layers.
Assumed oscillation parameters are
(\(\Delta m^2=2.5\times10^{-3}~\mbox{eV}^2\), 
\(\sin^2\theta_{23}=0.5\), 
\(\sin^2\theta_{13}=0.04\)
).
The top figure assumes neutrino oscillation in vacuum and the bottom
figure takes into account Earth's matter effect.
In the bottom figure, three high probability (\(> 40\%\)) regions 
are shown which correspond to the MSW resonance at 3 GeV in the core layer, 
the MSW resonance at 7 GeV in the mantle layer,
and the enhancement due to 
the core-mantle transition interference
\cite{Chizhov:1998ug} at the energy between the two MSW regions.
}
\label{fig:nue-flux}
\end{center}
\end{figure}
Fig. \ref{fig:nue-flux} shows the transition probability of
\(\nu_e\) to $\nu_\mu$
after traversing the Earth.
Note that for normal mass hierarchy,
the \(\nu_e\) flux is resonantly enhanced, and there is
no enhancement for anti-neutrinos; the situation is reversed for the
inverted mass hierarchy.  
In matter, the solar and the interference terms 
modify the $\nu_e$ enhancement in the resonance region by less than 5\%,
justifying our assumption of $\Delta m^2_{12}=0$.




\section{Data sample}









Super-Kamiokande is a 22.5~kt fiducial mass water Cherenkov detector
located at a depth of 2700m water equivalent in the Kamioka mine, Gifu, Japan. 
The detector is optically separated into two concentric cylindrical regions. 
The inner detector (ID) is instrumented with 11,146 20-inch 
photomultiplier tubes (PMT). 
The outer detector (OD) is instrumented with 1,885 8-inch PMTs.
Details of the detector can be found in Ref. \cite{Fukuda:2002uc}.
Physical quantities associated with 
a neutrino event such as the interaction vertex, the number of 
Cherenkov rings, the direction of each ring, particle identification (PID),
momentum, and number of Michel electrons are reconstructed by using 
hit timing and charge distributions of Cherenkov ring images recorded by
PMTs on the ID wall.

Atmospheric neutrino data are categorized into fully-contained (FC), 
partially-contained (PC), and upward-going muons (UP$\mu$).
In the FC events, all of their Cherenkov light is deposited in the ID.
In the PC events,
there is an exiting particle that deposits visible energy in the OD.
Neutrino interactions in the rock below the detector 
produce UP$\mu$ events for which 
muons either stop in the detector, or pass through the detector.
In the present analysis, we use 1489 live-days of FC, PC and 
1646 days UP$\mu$ neutrino data taken from May 1996 through
July 2001 during the Super-Kamiokande I period.

We divide the FC sample into sub-GeV and multi-GeV subsamples
according to the visible energy as
$E_{vis}<1.33$ GeV for sub-GeV, $E_{vis}>1.33$ GeV for multi-GeV
where the visible energy of an event is the total energy assuming all
Cherenkov light is from electromagnetic showers.
The sample is also divided into
single-ring and multi-ring by number of reconstructed Cherenkov rings, and into
$e$-like and $\mu$-like by PID of the most energetic ring.
Sub- and multi-GeV, $e$- and $\mu$-like events from the single-ring sample,
and sub- and multi-GeV $\mu$-like events from the multi-ring sample are used in
the analysis, as for the $\nu_\mu\rightarrow\nu_\tau$ two-flavor
oscillation analysis \cite{Ashie:2005ik}. 
We also divide the PC sample into 
`OD stopping events' and `OD through-going events' according
to energy
deposited in the OD \cite{Ashie:2004mr}. 
Finally, the UP$\mu$ sample is divided into upward stopping muons
(entering and stopping in the tank) and upward through-going muons
(passing through the tank).

\begin{figure*}
\includegraphics[width=150mm]{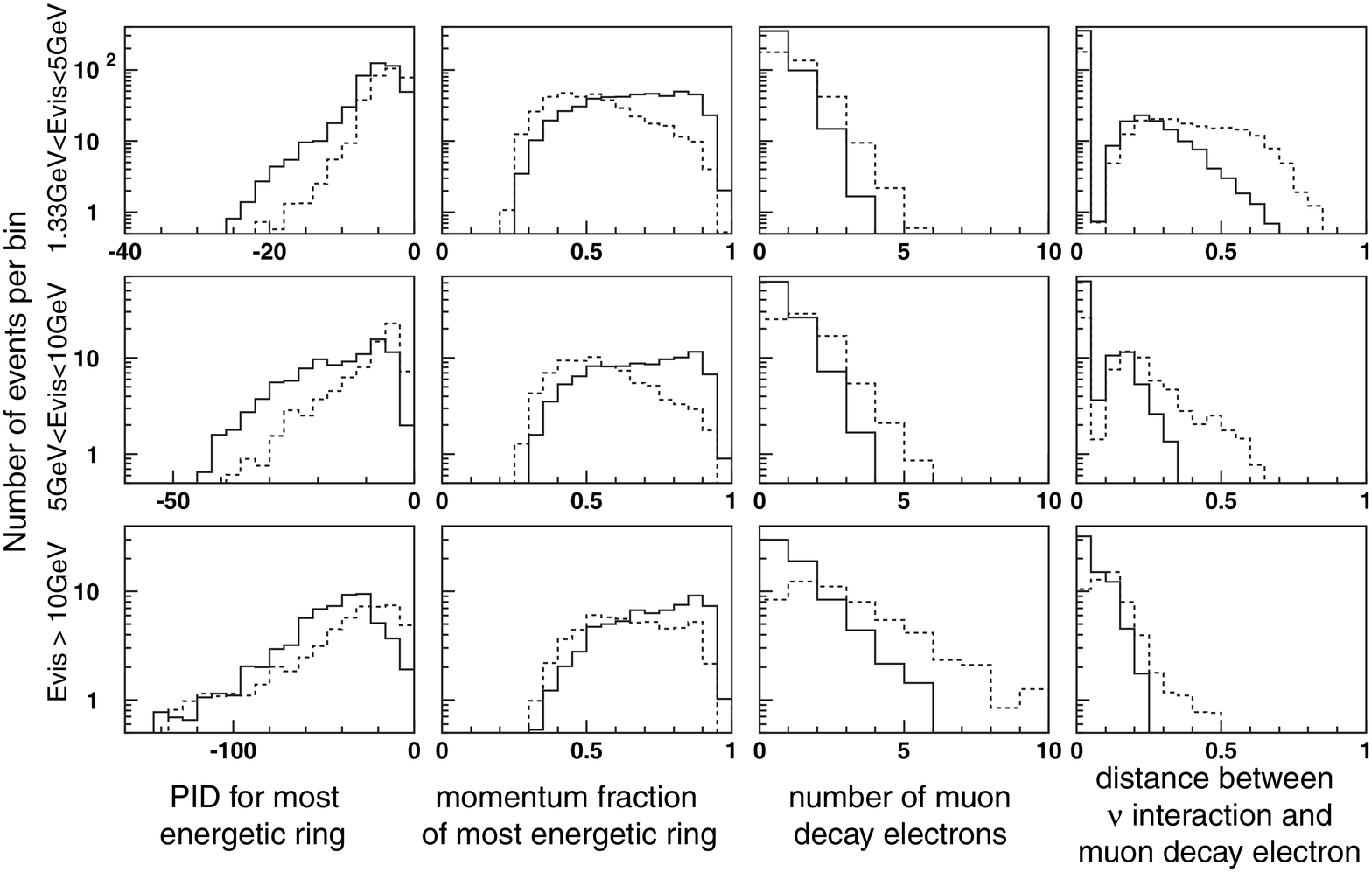}
\caption{Livetime normalized MC
distributions used in the $\nu_e$-enriched likelihood selection
for multi-GeV multi-ring $e$-like events. Plots correspond to $1.33<E_{vis}<5$
GeV, $5<E_{vis}<10$ GeV, and $E_{vis}>10$ GeV from top to bottom.
From left to right are shown
PID likelihood for the most energetic ring (more negative
means more electron-like), 
momentum fraction of the most energetic ring, number of Michel decay electrons,
and square root of distance between decay electron and neutrino interaction vertex divided 
by visible energy of most energetic ring in (cm/MeV)$^{1/2}$.
Solid histograms are $\nu_e+\bar{\nu}_e$ CC, 
dashed are backgrounds (NC + $\nu_\mu$ CC + $\bar{\nu}_\mu$ CC).
The \(\nu_e\) or \(\bar{\nu}_e\) CC signals tend to give more electron-like PID
and higher momentum fraction.
In contrast, backgrounds tend to give more muon decay electrons
and the longer decay electron distance due to energetic muons produced by
\(\nu_\mu\) or \(\bar{\nu}_\mu\) CC interactions.
}
\label{likelihood}
\end{figure*}

As mentioned in the previous section, an excess of upward-going 
$\nu_e$ and/or $\bar{\nu}_e$ in the 
several GeV region is expected for certain 
oscillation parameter sets. To improve the sensitivity to this case,
a $\nu_e$-enriched sample is selected from the multi-GeV multi-ring $e$-like
sample and is used in addition to the standard oscillation analysis samples.
The $\nu_e$-enriched selection is based on a likelihood analysis
using PID likelihood, momentum fraction of the most energetic ring, number of 
muon decay electrons, 
and distance between muon decay electron and primary neutrino
interaction
position.
Figure \ref{likelihood} shows distributions of Monte Carlo (MC) events
used to obtain the $\nu_e$-enriched likelihood function.
The $\nu_e+\bar{\nu}_e$ charged current (CC) 
and $\nu_\mu$ CC + $\bar{\nu}_\mu$ CC +
neutral current (NC) interactions are separately shown.
These distributions are divided into 5 energy regions;
1.33-2.5 GeV, 2.5-5 GeV, 5-10 GeV, 10-20 GeV, and \(>20\) GeV.
Each normalized signal (background) distributions forms 
a probability density function
\(PDF^{Sig}_{i,j}(x_i)\) (\(PDF^{BG}_{i,j}(x_i)\)),
where \(i\) denotes the four observed variables and \(j\) is
five energy bins.
Then the \(\nu_e\)-enriched selection criterion
for each event 
is defined as
\(\sum_{i=1}^{4} 
\{\log(PDF^{Sig}_{i,j}(x_i))-\log(PDF^{BG}_{i,j}(x_i))\} > 0\).
\begin{table}[htb]
\caption{ \label{efftable}
Breakdown of multi-GeV multi-ring $e$-like events in the MC sample
before and after a $\nu_e$-enrichment based on
a likelihood analysis.
For each interaction mode, the number of events normalized to 1489
live-days and the fraction
are shown assuming pure $\nu_\mu \leftrightarrow \nu_\tau\) 
two-flavor oscillation
(\(\Delta m^2=2.5\times10^{-3} \mbox{eV}^2\), 
\(\sin^2\theta_{23}=0.5\), 
\(\sin^2\theta_{13}=0\)
).
Survival efficiency for $\nu_e+\bar{\nu}_e$ CC
events is 66.8\%. }
\begin{tabular}{ccccc}
\hline
\hline
multi-GeV &&&& \\
multi-ring& ~$\nu_e+\bar{\nu}_e$ CC~ & ~$\nu_\mu+\bar{\nu}_\mu$
 CC~ & ~NC~ & ~~total~~\\
$e$-like events&&&& \\
\hline
no likelihood cut & 630.8 & 242.2 & 258.0 & 1131.1 \\
 & 55.8\% & 21.4\% & 22.8\% & 100\%\\
likelihood cut & 421.3 & 49.1 & 86.8 & 557.3 \\
 & 75.6\% & 8.8\% & 15.6\% & 100\% \\
\hline
efficiency & 66.8\% & & & \\
\hline
\hline
\end{tabular}
\end{table}
As is shown in Table \ref{efftable},
the $\nu_e+\bar{\nu}_e$ fraction in the multi-GeV 
multi-ring $e$-like sample is improved to 
75.6\% after the likelihood cut. 
A summary of the number of observed and expected 
FC multi-ring $e$-like events
is shown in Table \ref{tb:eventnumber1} in the appendix.


\section{\label{sec:analysis}oscillation analysis} 

%
%
%
%
%
%
%

The oscillation analysis is performed by comparing data with MC
equivalent to 100 years of detector exposure.
The atmospheric neutrino flux calculation from \cite{Honda:2004yz} and 
neutrino interaction model (NEUT) \cite{Ashie:2005ik, Hayato:2002sd} 
are used to simulate interactions
with the nuclei of water, or in the case of upward muons,
the nuclei of the rock surrounding the detector. 
We use a GEANT-based full detector simulation
to generate the MC neutrino events.

We employ a $\chi^2$ test to perform three-flavor oscillation analysis.
All events are divided into 37 momentum or energy bins 
(10$+$5~bins for FC single- and multi-ring $e$-like, 
8$+$4~bins for FC single- and multi-ring $\mu$-like, 
4$+$4~bins for OD stopping and OD through-going PC, 
and 1$+$1~bin for upward stopping and through-going muons).
The reconstructed quantities used 
for momentum binning for the various event classes are:
the observed momentum of the charged lepton for 
FC single-ring $e$- and $\mu$-like events ($P_{lep}$),
the sum of the energies of the observed rings (reconstructed particles) 
considering
particle mass for FC multi-ring $e$-like events ($E_{tot}$), and 
the sum of visible energies of observed rings for 
FC multi-ring $\mu$-like and PC events ($E_{vis}$).
Note that the binning of the energy scale differs from that
of \cite{Ashie:2004mr,Ashie:2005ik}, i.e. 
events in the multi-GeV energy range are divided more finely
in order to obtain better sensitivity to $\Delta m^2$.
Each momentum bin is also divided into 10 bins equally spaced between 
$\cos\Theta=-1$ and $\cos\Theta=+1$ 
($-1<\cos\Theta<0$ for UP$\mu$ events), 
where $\cos\Theta$ is the cosine of the zenith angle of 
the reconstructed particle direction.
Table~\ref{tb:eventnumber1} summarizes the number of observed and 
expected FC, PC and UP\(\mu\) events for each bin. 
The total number of bins is 370.
The number of events in each bin is compared with expectation 
and a $\chi^2$ value is calculated according to a
Poisson probability distribution defined by the 
following expression:

 \begin{eqnarray}
 \chi^2 &=& \sum_{n=1}^{370} \left[ 2 \left\{ N_{exp}^n \left( 
 1+\sum_{i=1}^{45} f_i^n \cdot
 \epsilon_i \right) - N_{obs}^n \right\} \right. \nonumber \\
  && \left. + 2 N_{obs}^n \ln \left( \frac{N_{obs}^n}{N_{exp}^n \left( 1+ 
 \sum_{i=1}^{45} f_i^n \cdot \epsilon_i \right) } \right)  \right] \nonumber \ \\
  && + \sum_{i=1}^{43} \left( \frac{\epsilon_i}{\sigma_i} \right)^2
 \label{eq:chisq}
 \end{eqnarray}

 \noindent 
 where
 
 \begin{center}
 \begin{tabular}{ll}
 $N_{obs}^n$ & Number of observed events in $n$-th bin \\
 $N_{exp}^n$ & Number of expected events in $n$-th bin \\
 $\epsilon_i$ & $i$-th systematic error term \\
 $f_i^n$      &  Systematic error coefficient \\
 $\sigma_i$  & 1 sigma value of systematic error 
 \end{tabular}
 \end{center}

The expectation $N_{exp}^n$ is calculated
using MC events corrected by oscillation probability. 
Here, systematic uncertainty factors explicitly multiply
the $N_{exp}^n$.
We considered 45 systematic error sources, which come from
detector calibration,
neutrino flux, neutrino interactions and event selection.
Most of them are in common with those listed in \cite{Ashie:2005ik},
and additional systematic uncertainties 
related to backgrounds of the $e$-like sample and upward-going muons,
and sample normalization of $e$-like events
are estimated as listed in Table \ref{tab:syserror}.
\begin{table}[htb]
\caption{Sources of systematic errors in addition to those
in common with the $\nu_\mu \leftrightarrow \nu_\tau$ oscillation analysis 
\cite{Ashie:2005ik}.
Non-$\nu$ backgrounds in UP$\mu$ samples are treated as 
fitting parameters
in this analysis, while they are taken into account by modifying statistical
error size in Ref. \cite{Ashie:2005ik}.} 
\begin{tabular}{lc}
\hline
\hline
 & estimated error size (\%) \\
\hline
Non-$\nu$ background
\footnote{
Cosmic ray muon backgrounds are assigned as systematic errors
to the most horizontal zenith angle bins ($-0.1<\cos\Theta<0$).
} & \\
~~~~~~upward through-going muons & 3.0 \\
~~~~~~upward stopping muons & 17 \\
Non-($\nu_e$ CC) background 
\footnote{
Contamination of \(\nu_\mu\) CC interactions.
} & \\
~~~~~~multi-GeV single-ring $e$-like & 14 \\
~~~~~~multi-GeV multi-ring $e$-like & 20 \\
sample normalization & free \\
~~~of multi-GeV multi-ring $e$-like & \\
OD stopping PC/through-going PC & 12 \\
~~~separation & \\
\hline
\hline
\end{tabular}
\label{tab:syserror}
\end{table}
Among 45 errors, only 43 contribute to the \(\chi^2\), because the
overall normalization and sample normalization of multi-GeV multi-ring
\(e\)-like are allowed to be free.
The $f_i^n$ values are calculated and tabulated in advance for every bin and 
for every systematic error source.
A global scan is carried out on a 
($\log_{10}(\Delta m^2)$, $\sin^2\theta_{23}$, $\sin^2\theta_{13}$) grid
minimizing $\chi^2$ with respect to 45 systematic error parameters.
The $\epsilon_i$ values are fit in order 
to minimize the $\chi^2$ value.
We used $\epsilon_i$ such that the first derivative of $\chi^2$ with
respect to $\epsilon_i$ is zero 
($\frac{\delta\chi^2}{\delta\epsilon_i}=0$), 
which can be obtained by solving linear equations~\cite{Fogli:2002pt}.
Since this equation has non-linear terms for our $\chi^2$ definition, 
we use an approximate solution obtained by an iteration method.

\begin{figure}[ht]
\begin{minipage}{80mm}
\includegraphics[width=80mm,height=110mm]{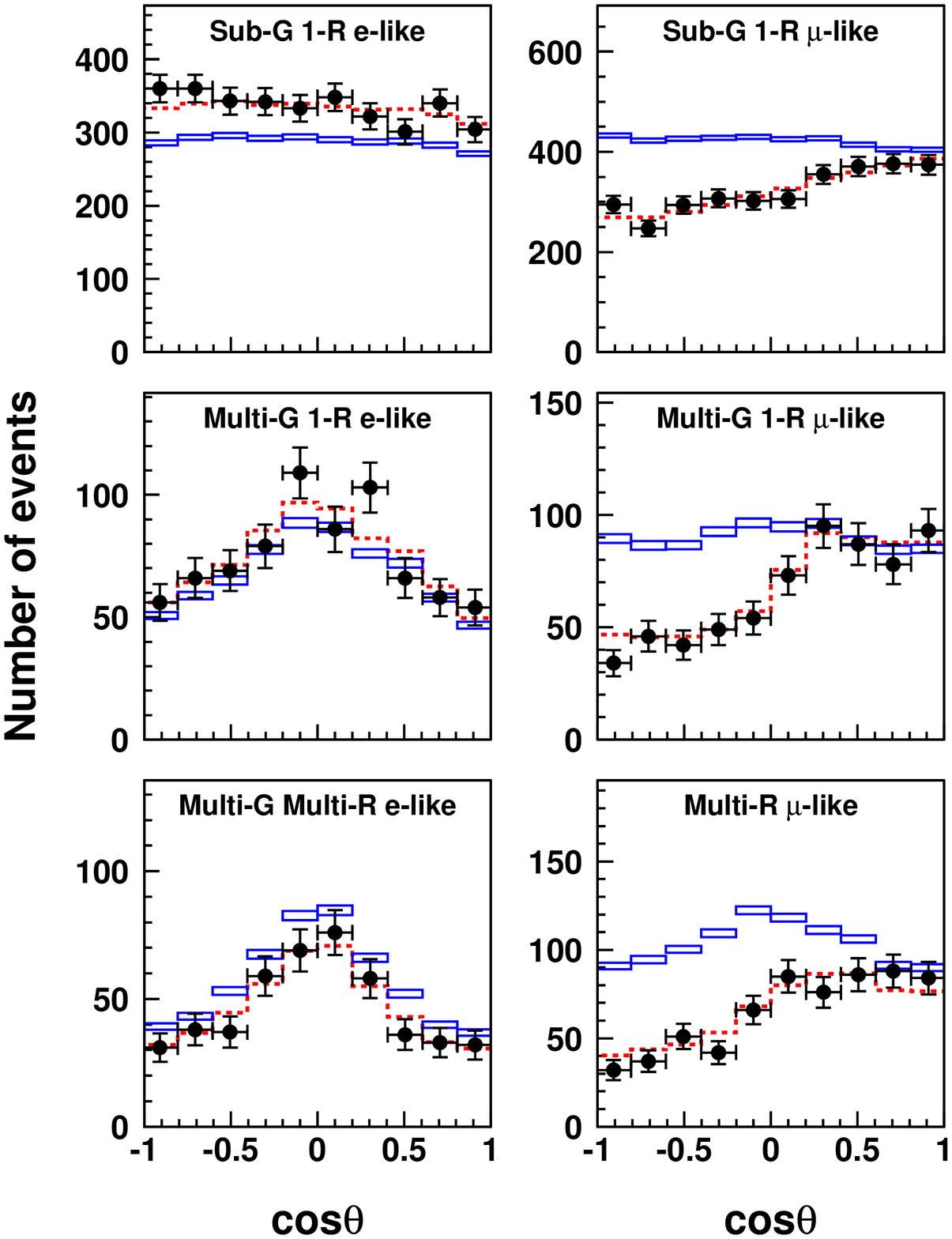}
\end{minipage}
\begin{minipage}{80mm}
\includegraphics[width=80mm]{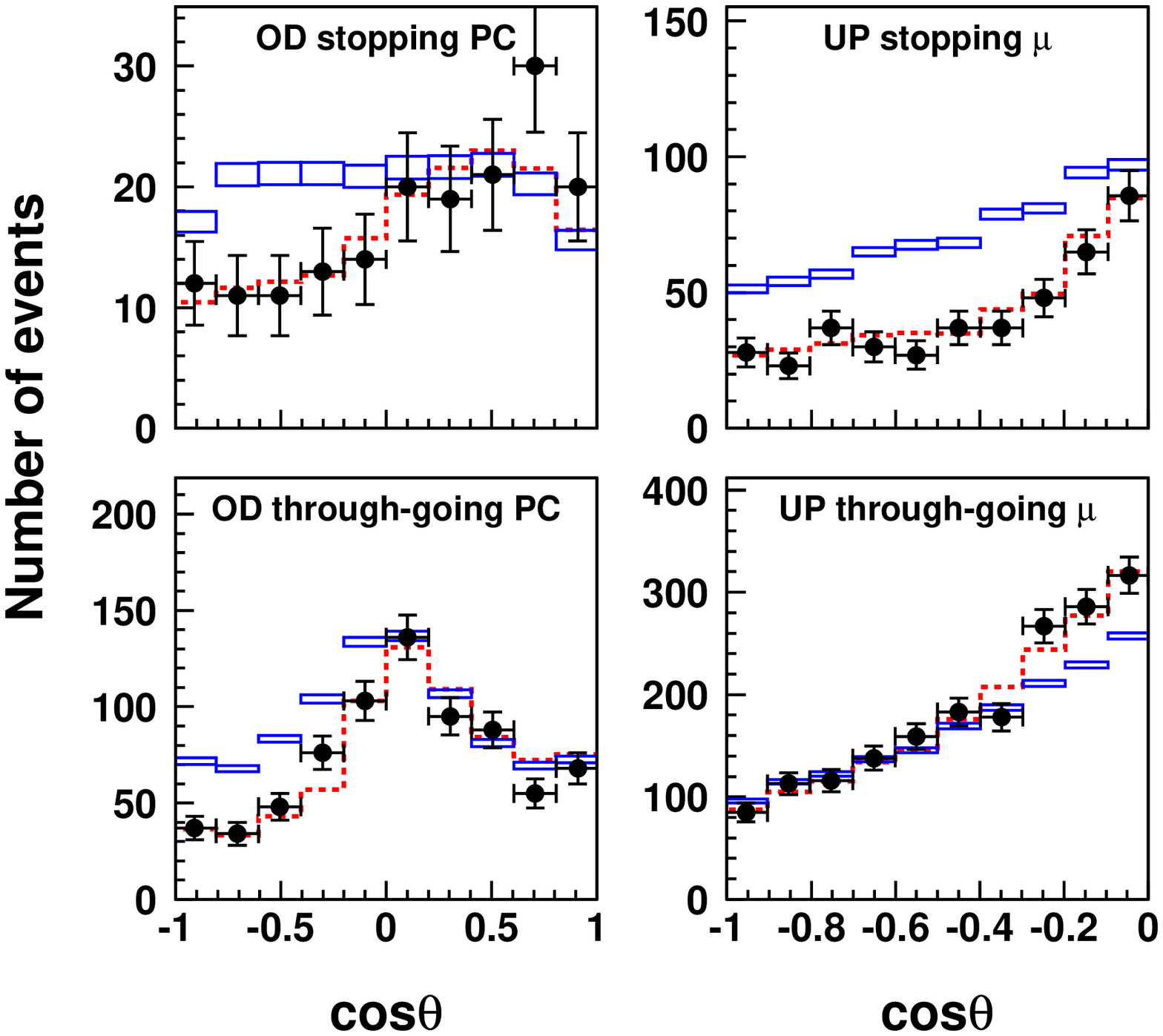}
\end{minipage}
\caption{
Zenith angle distributions of
FC $e$-like, $\mu$-like, PC, and
UP$\mu$ are shown for data (filled circles with statistical error bars),
MC distributions without oscillation (boxes) and
best-fit distributions (dashed).
The non-oscillated MC shows the distribution without fitting
and the box height shows the statistical error.
In the case of non-zero \(\theta_{13}\), 
matter enhanced excess of electron-like events
is expected in the zenith angle of \(-1 < \cos\theta < -0.2\) regions
in the multi-GeV 1-ring and multi-ring electron-like samples.
The \(\nu_\mu\) in the resonance regions populate mainly in 
the multi-GeV single-ring muon, multi-ring muon, 
two PC, and UP stopping \(\mu\) samples.
}
\label{fig:zenith}
\end{figure}

\begin{figure}[ht]
\begin{center}
\includegraphics[width=80mm]{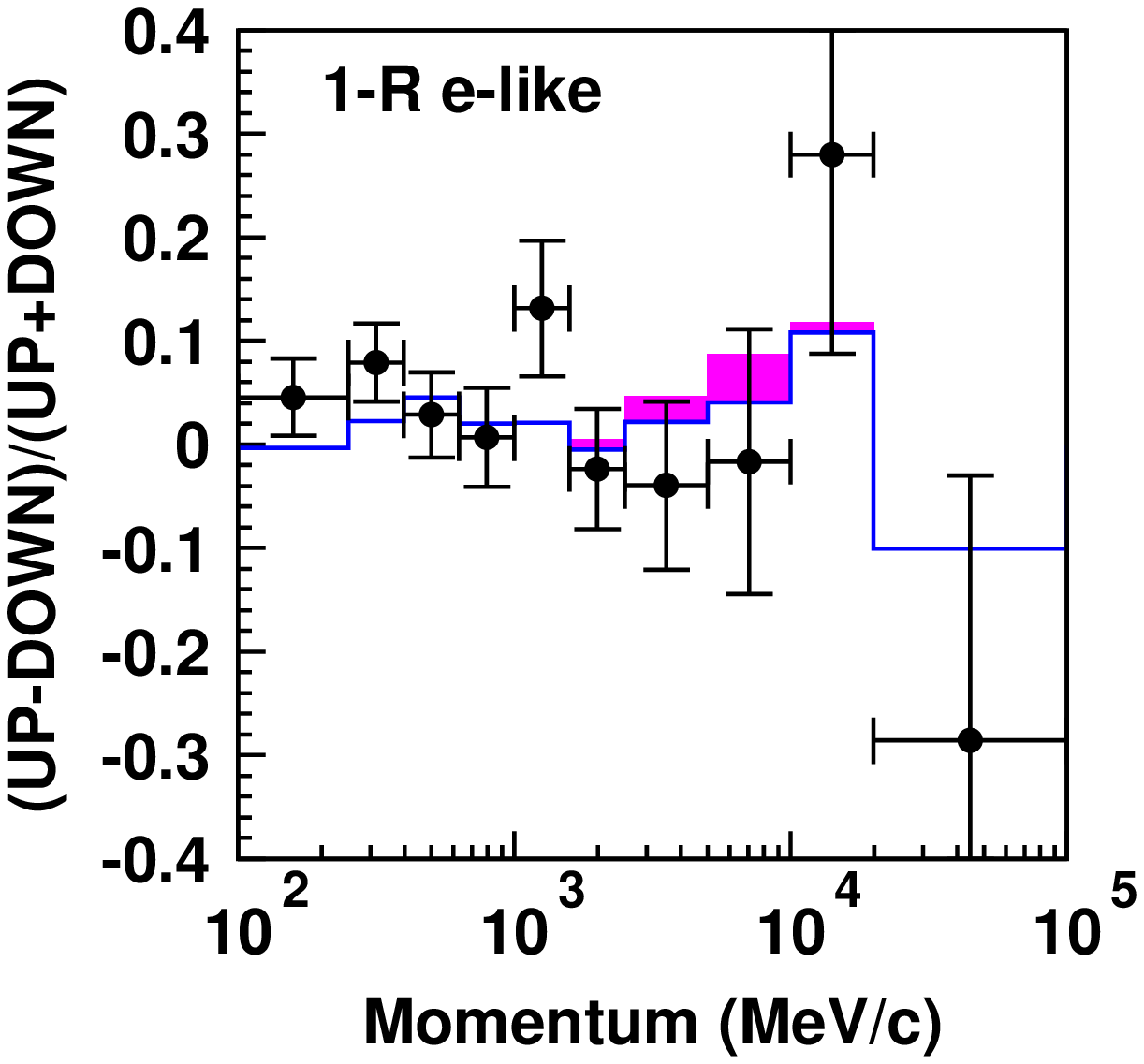}
\vspace*{3mm}
\includegraphics[width=80mm]{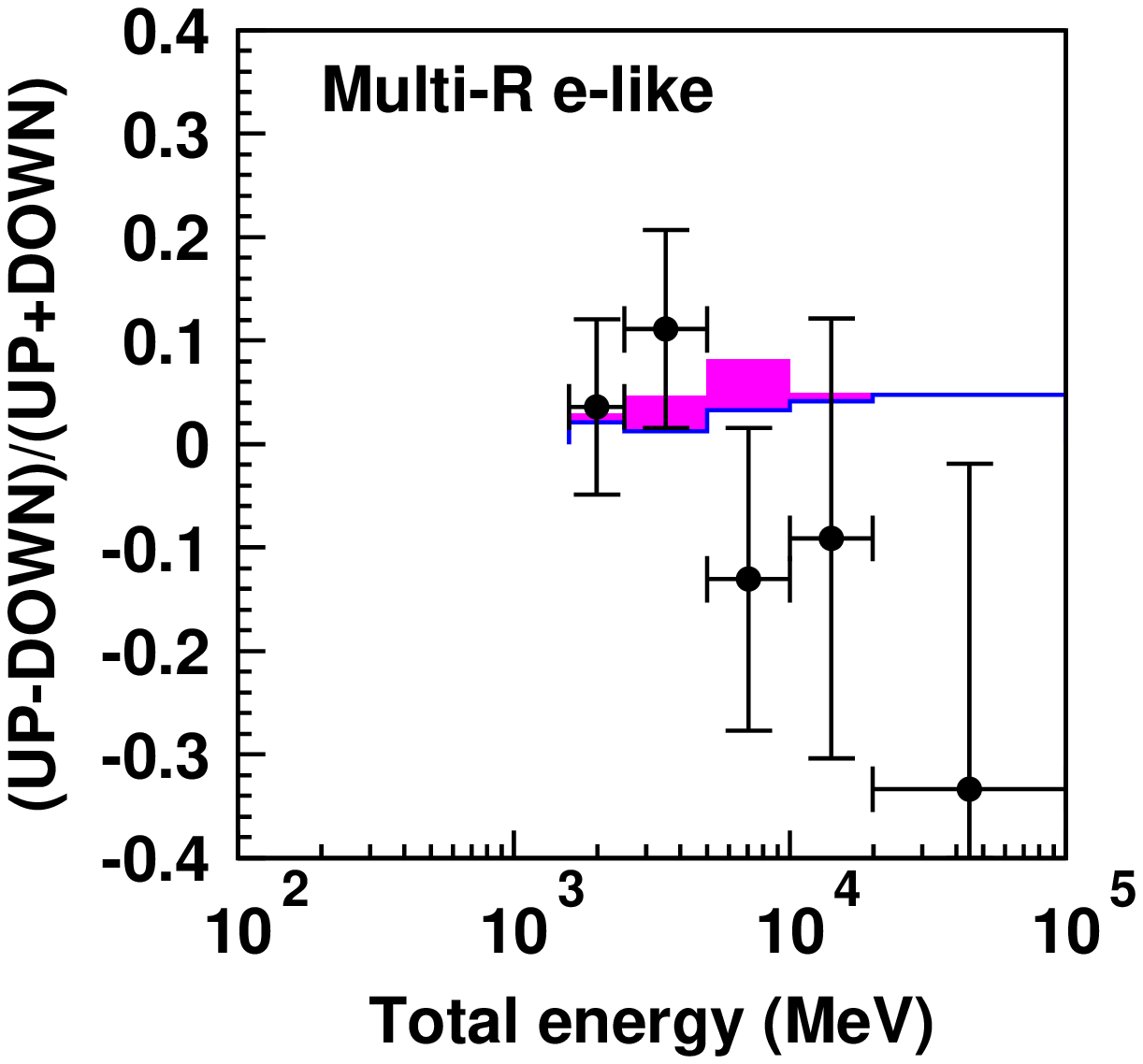}
\caption{
Asymmetry (UP-DOWN)/(UP+DOWN) as a function of particle momentum
for FC single-ring $e$-like events (top) and 
as a function of total energy for FC multi-ring $e$-like events (bottom),
where UP (DOWN) refers to the number of events in
$-1.0$$<$$\cos\Theta$$<$$-0.2$ ($0.2$$<$$\cos\Theta$$<$$1.0$).
Filled circles represent data (error bars are statistical), the
line represents best-fit distributions under a
normal hierarchy assumption, and the
filled area on the best-fit line represents 
the expected excess due to matter effect for 
($\Delta m^2$, $\sin^2\theta_{23}$, $\sin^2\theta_{13}$)$=$
(2.5$\times$10$^{-3}$~eV$^2$, 0.5, 0.04).
}
\label{fig:mme_updown}
\end{center}
\end{figure}

\begin{figure}[ht]
\begin{center}
\begin{minipage}{70mm}
\includegraphics[width=70mm]{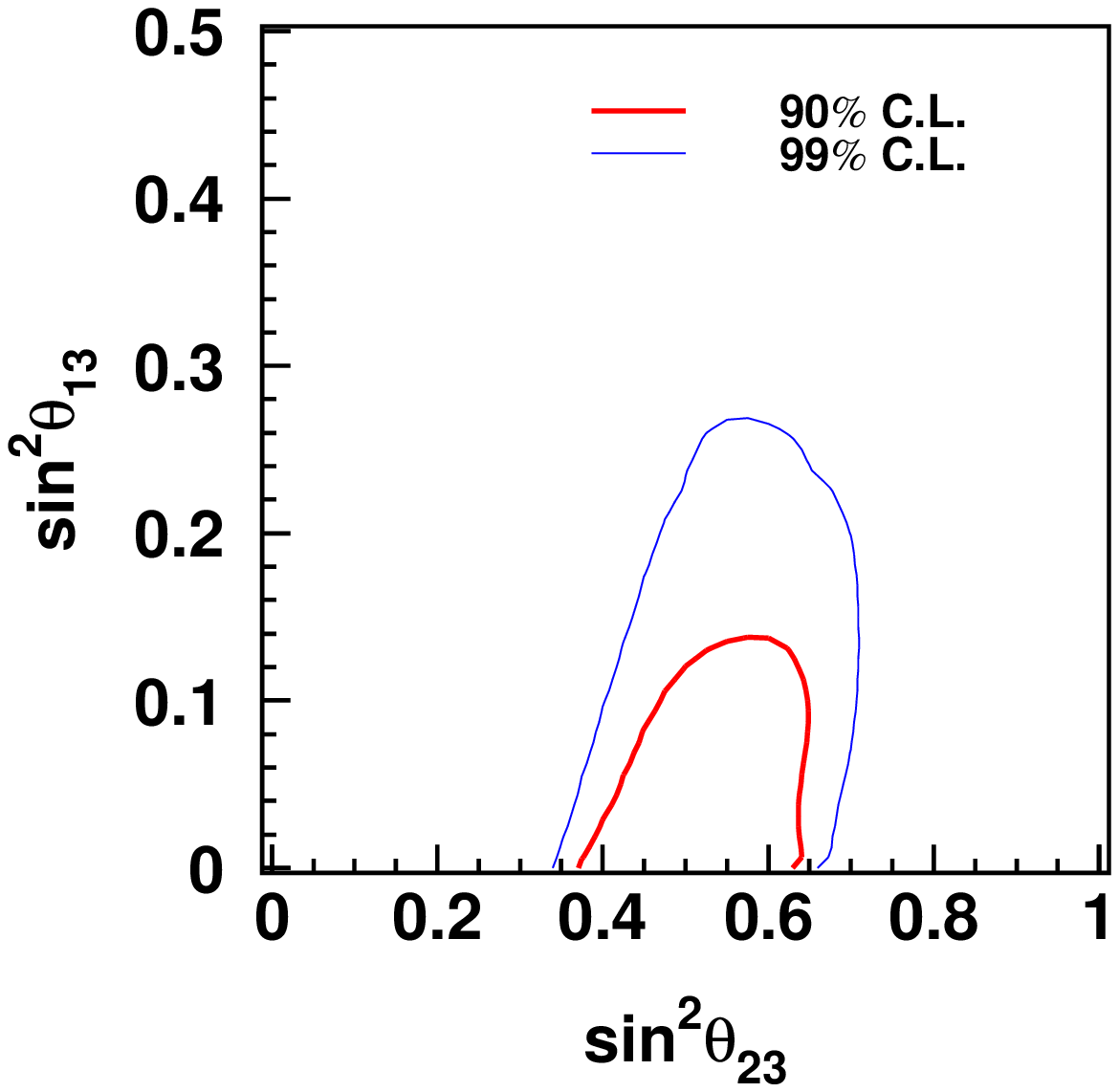}
\end{minipage}
\begin{minipage}{70mm}
\includegraphics[width=70mm]{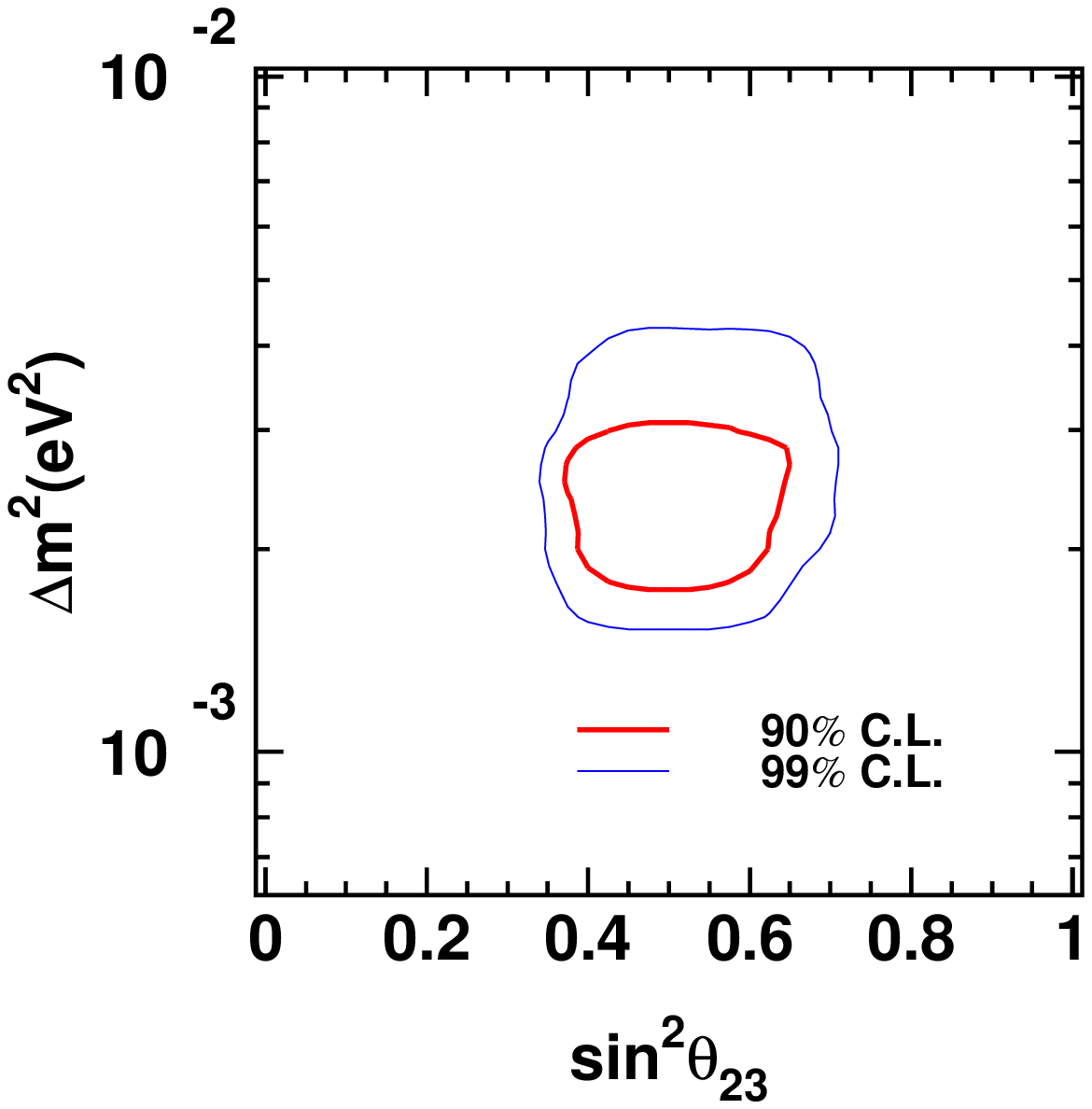}
\end{minipage}
\begin{minipage}{70mm}
\includegraphics[width=70mm]{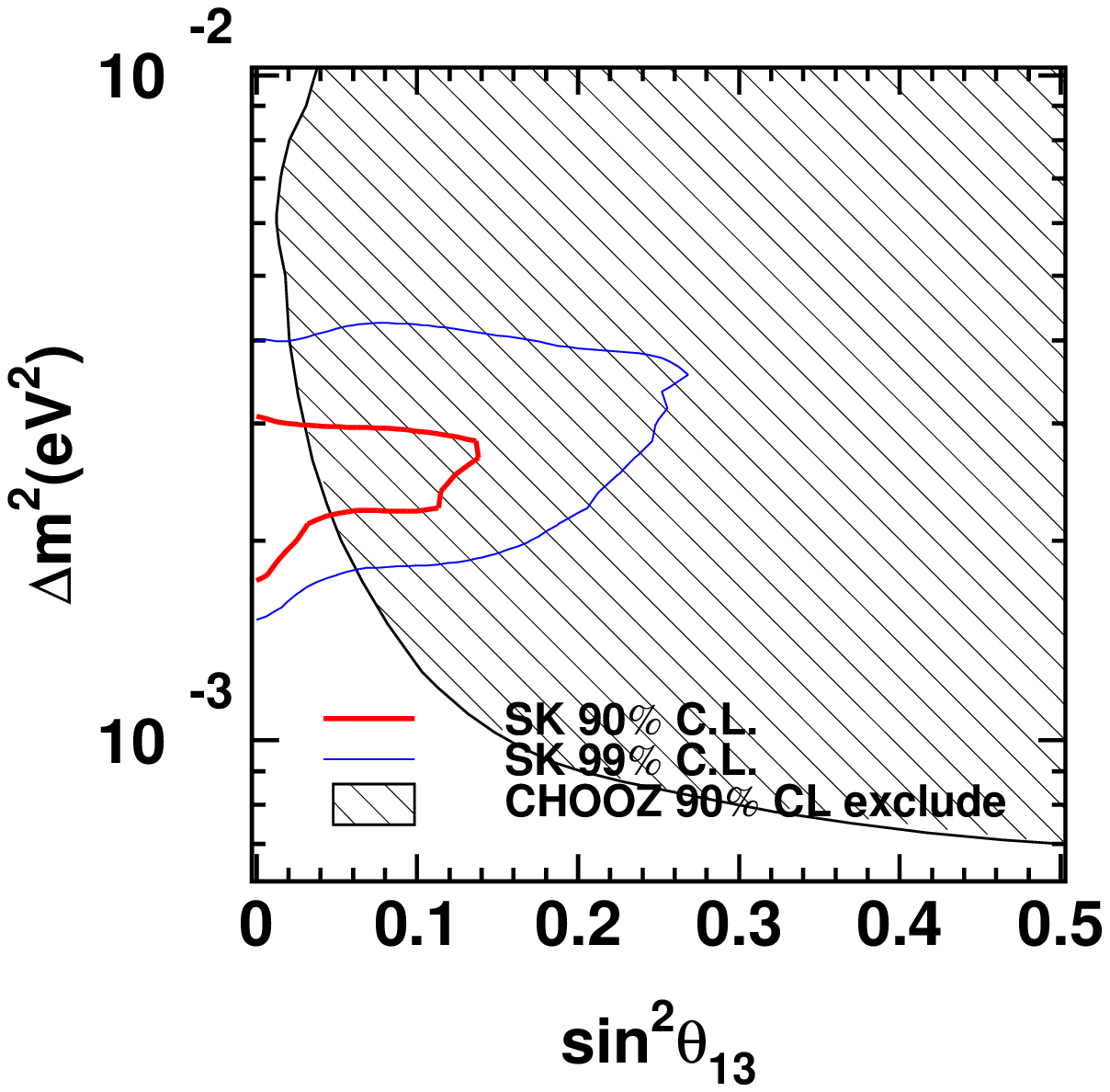}
\end{minipage}
\caption{
90~\% (thick line) and 99~\% (thin line) confidence level
allowed regions are
shown in $\sin^2\theta_{13}$ vs $\sin^2\theta_{23}$ (top),
$\Delta m^2$ vs $\sin^2\theta_{23}$ (middle),
and $\Delta m^2$ vs $\sin^2\theta_{13}$ (bottom).
Normal mass hierarchy ($\Delta m^2$$>$0) is assumed.
The shaded area in the bottom figure shows the region excluded
by the CHOOZ reactor neutrino experiment.
}
\label{fig:allowed}
\end{center}
\end{figure}

A global scan of the oscillation parameter grid
assuming normal mass hierarchy results in
a minimum $\chi^2$ value of $\chi^2_{min}$$=$377.39/368~DOF 
at the grid point 
($\Delta m^2$, $\sin^2\theta_{23}$, $\sin^2\theta_{13}$)$=$
(2.5$\times$10$^{-3}$~eV$^2$, 0.5, 0.0), which is consistent with
$\nu_\mu$ $\leftrightarrow$ $\nu_\tau$ two-flavor oscillation.
Figure \ref{fig:zenith} shows the zenith angle distribution
of each data sample
overlaid with non-oscillated and best-fit expectations.
The fitted distributions agree well with data.
Figure \ref{fig:mme_updown} shows the up-down asymmetry
as a function of particle momentum or total energy.
The asymmetry (UP-DOWN)/(UP+DOWN) distributions are consistent
with the fitted expectation.
No significant excess due to matter effect is seen in the
upward-going multi-GeV $e$-like sample, suggesting no evidence
for non-zero \(\theta_{13}\).
Allowed regions of neutrino oscillation parameters are obtained
based on the \(\chi^2\) defined in Eq. \ref{eq:chisq}.
The 2-dimensional 
90~\% (99~\%) confidence level allowed region is defined to be
$\chi^2$$=$$\chi^2$$_{min}$$+$4.6~(9.2) and
obtained as shown in Fig.~\ref{fig:allowed}.
The region corresponding to $\sin^2\theta_{13}$$<$0.14 and
0.37$<$$\sin^2 \theta_{23}$$<$0.65
is allowed at 90\% confidence level.

\begin{figure}[ht]
\begin{center}
\begin{minipage}{70mm}
\includegraphics[width=70mm]{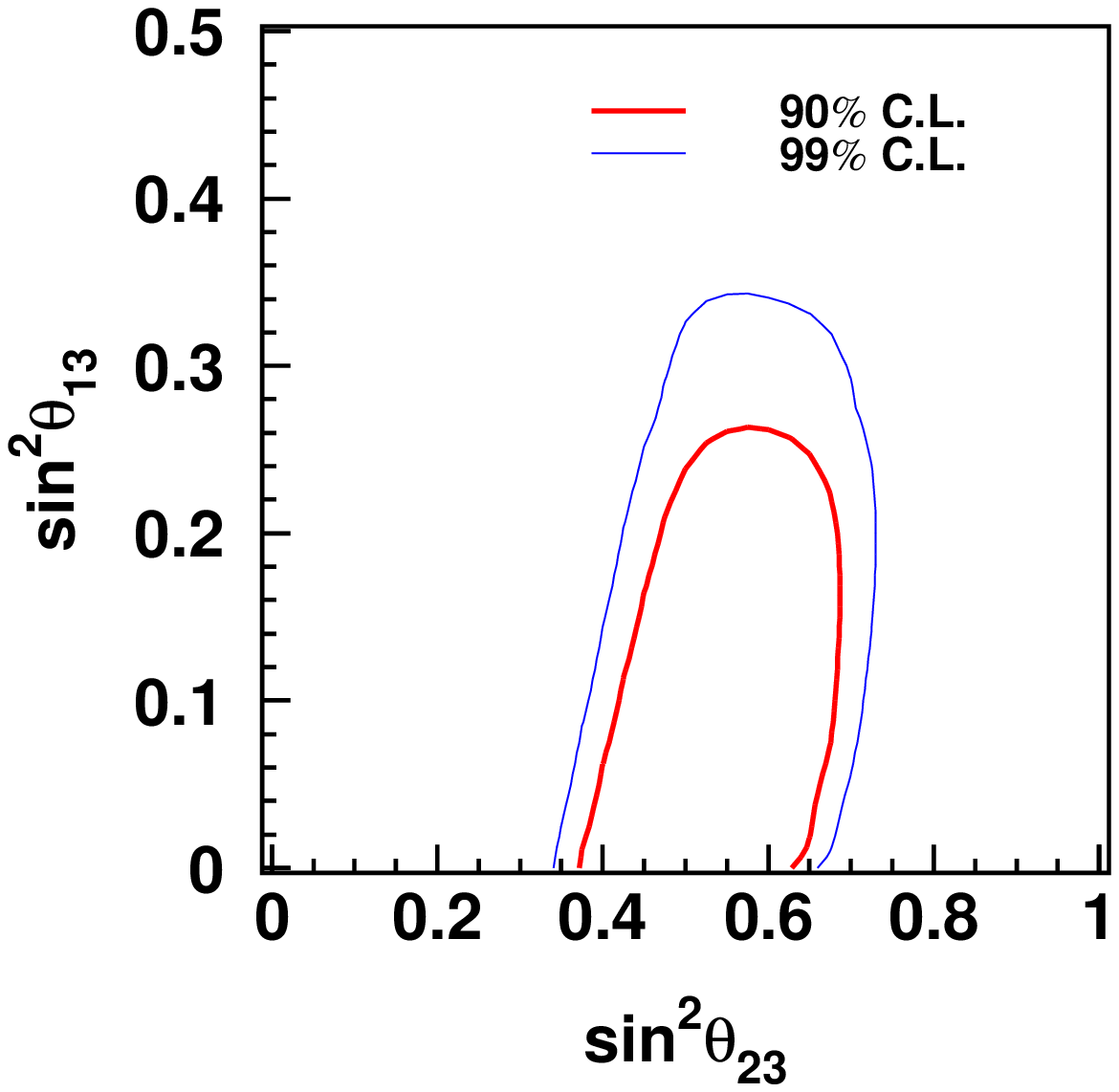}
\end{minipage}
\begin{minipage}{70mm}
\includegraphics[width=70mm]{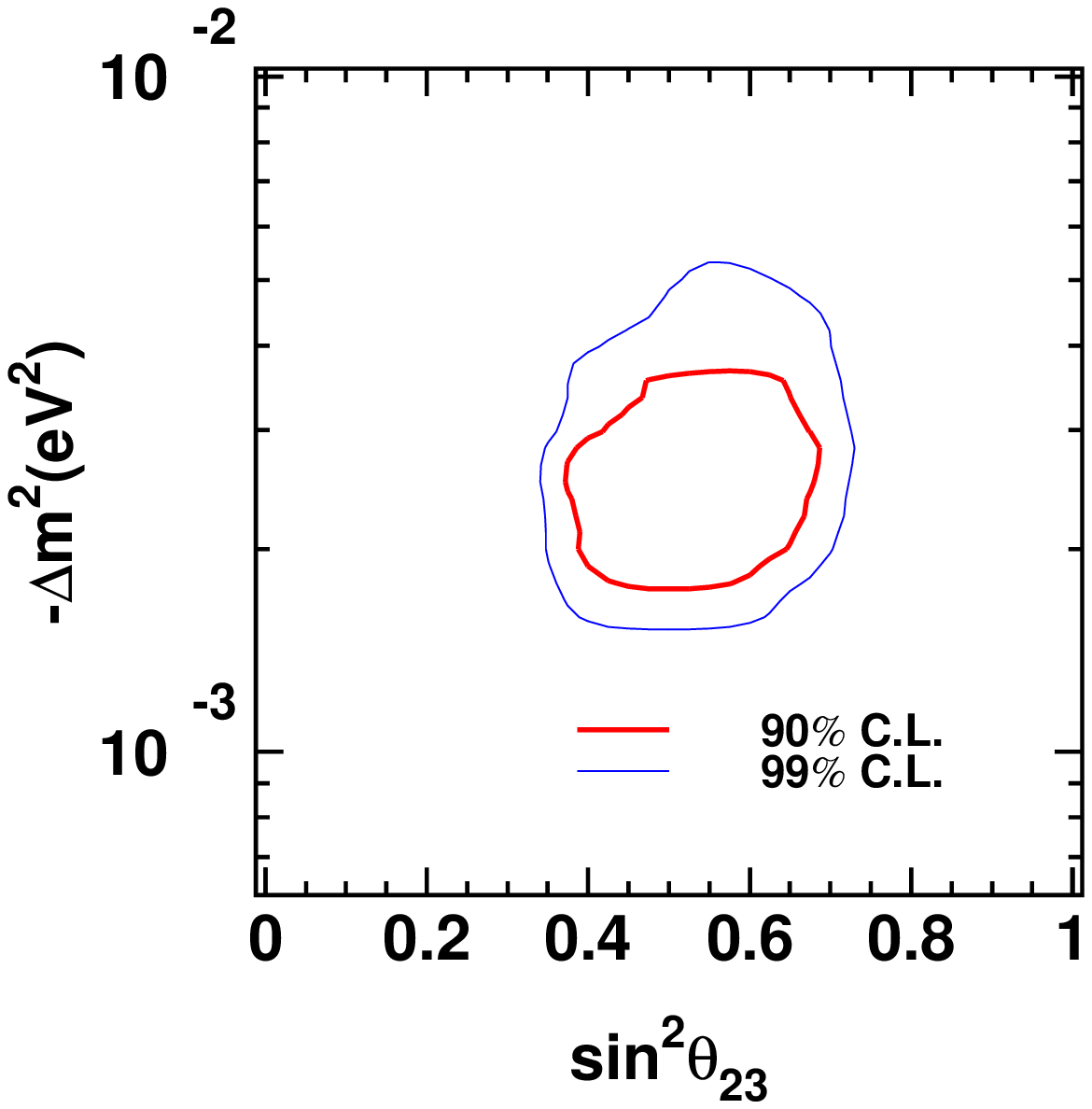}
\end{minipage}
\begin{minipage}{70mm}
\includegraphics[width=70mm]{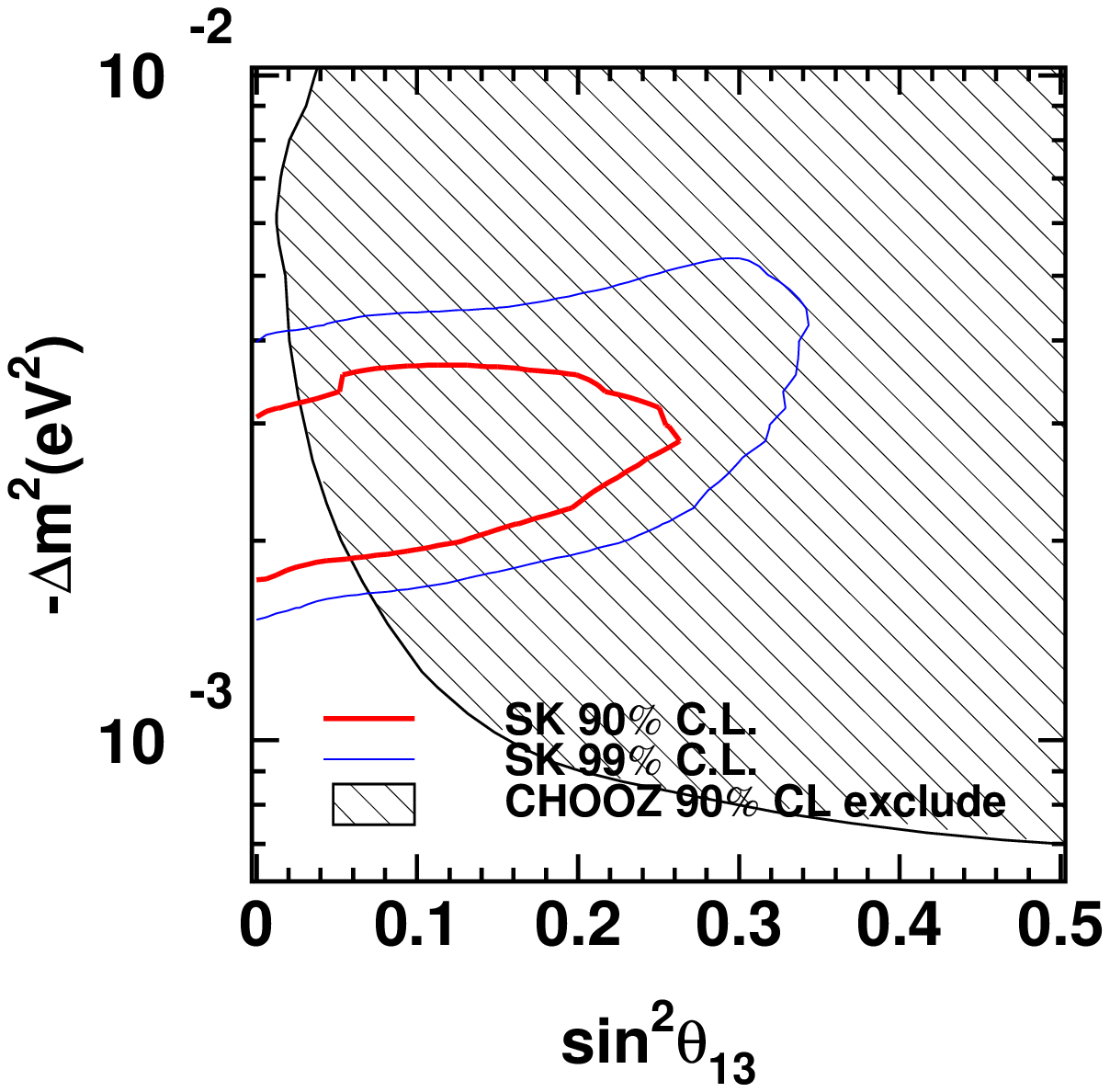}
\end{minipage}
\caption{
90~\% (thick line) and 99~\% (thin line) confidence level
allowed regions
assuming inverted mass hierarchy ($\Delta m^2$$<$0),
shown in the same orientation as in Fig.~\ref{fig:allowed}.
}
\label{fig:allowed_inv}
\end{center}
\end{figure}

\begin{figure}[ht]
\begin{center}
\includegraphics[width=80mm]{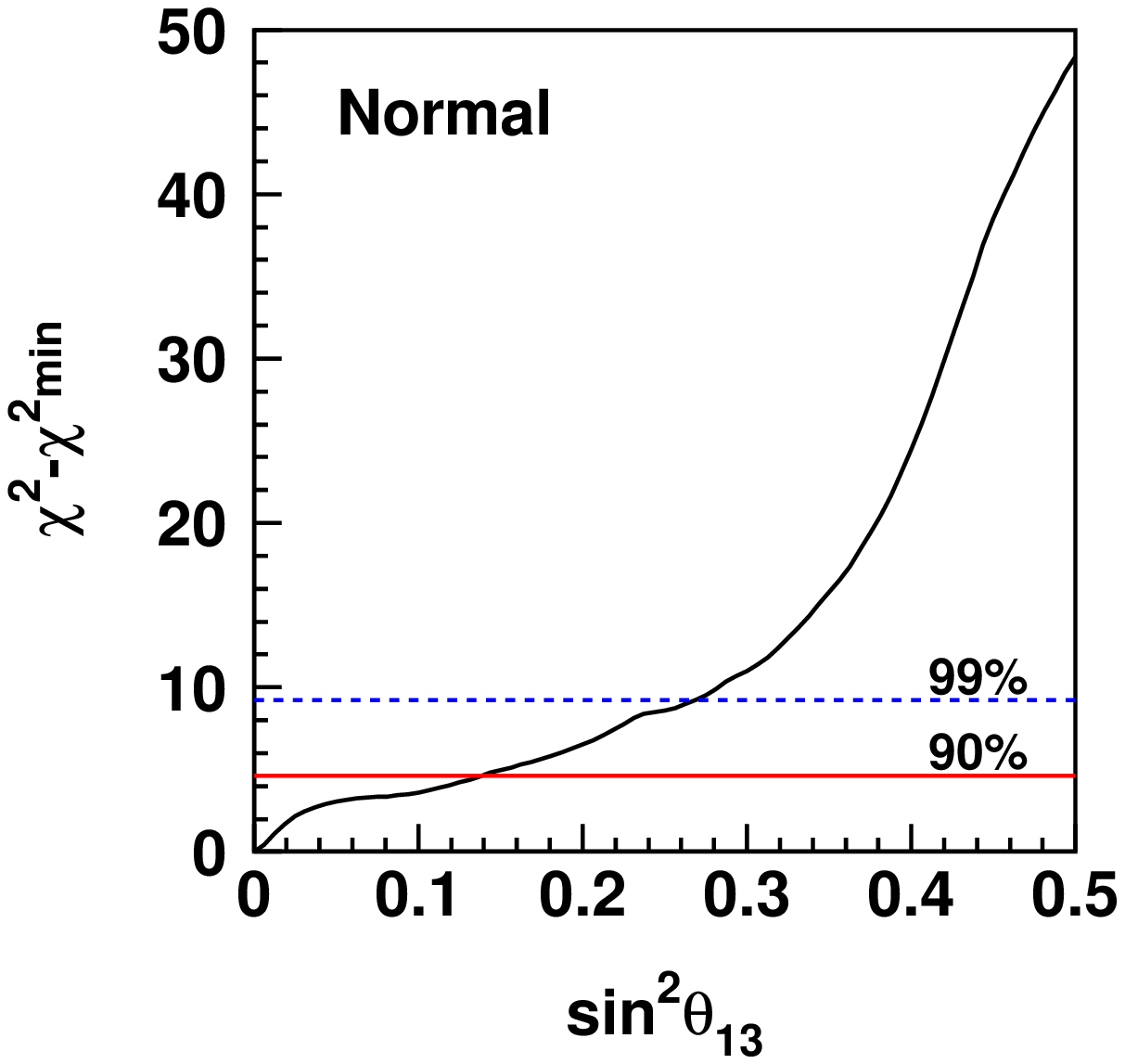}
\vspace*{3mm}
\includegraphics[width=80mm]{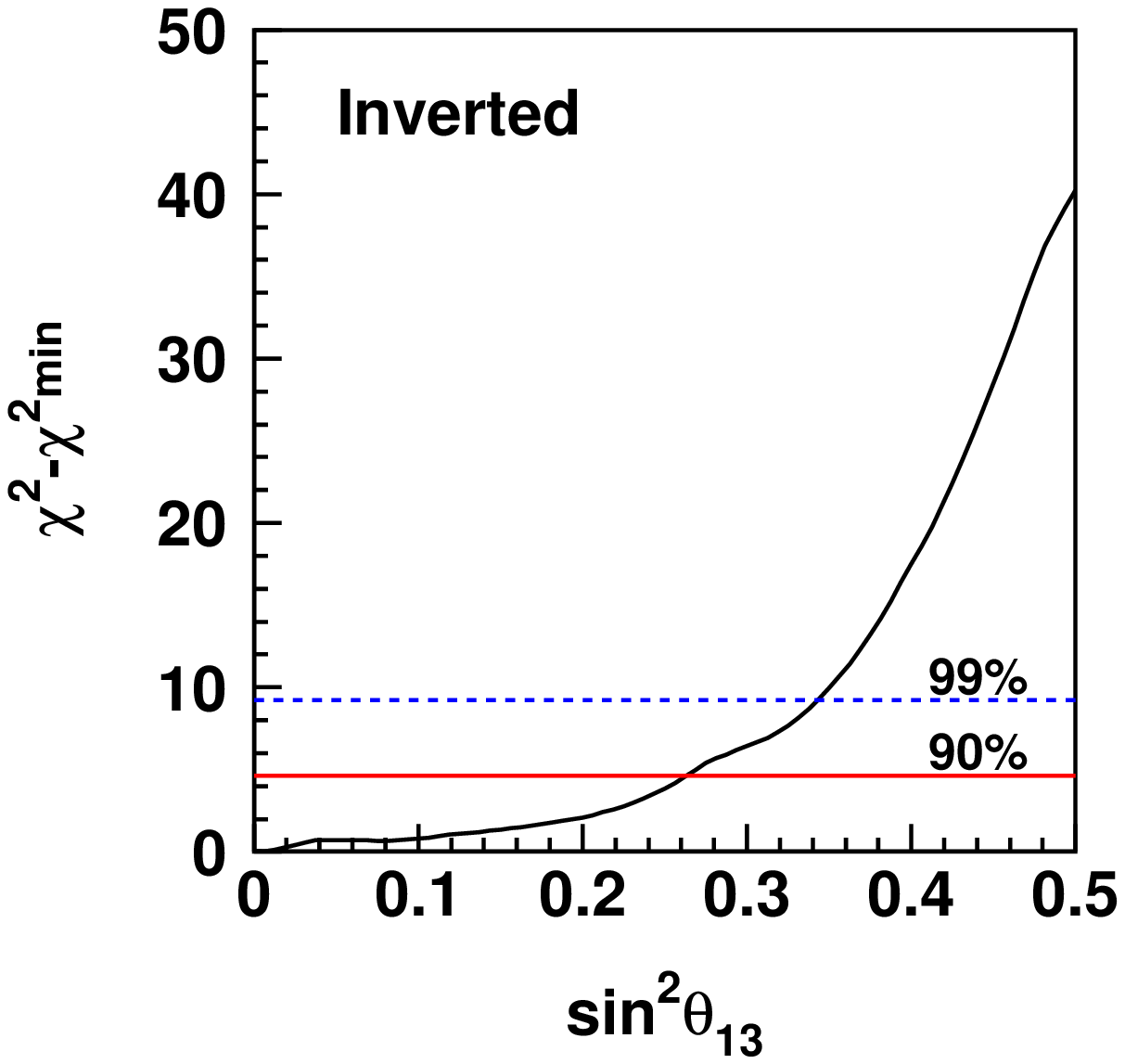}
\caption{
For normal (top) and inverted (bottom) mass hierarchy assumption,
$\chi^2-\chi^2_{min}$ are shown as a function of $\sin^2\theta_{13}$.
Minimum \(\chi^2\) values for each \(\Delta m^2\) and \(\sin^2\theta_{23}\)
are selected and plotted at each $\sin^2\theta_{13}$.
Solid (dashed) lines show 2-dimensional 
90 \% (99 \%) confidence level allowed regions.
}
\label{fig:sin2th13}
\end{center}
\end{figure}

Finally, we tested the inverted mass hierarchy hypothesis.
Water Cherenkov detectors, such as Super-K, cannot
discriminate between neutrinos and anti-neutrinos on an event-by-event basis.
However, mass hierarchy affects the expected number of $e$-like events. 
Because of the lower cross section of $\bar{\nu}_e$,
an enhancement of the multi-GeV $\nu_e$-rich sample is expected to be
suppressed and therefore the constraint on $\theta_{13}$ will be
weakened for the inverted mass hierarchy case.
Allowed regions assuming inverted mass hierarchy are also obtained
and shown in Fig.~\ref{fig:allowed_inv}.
$\chi^2$$_{min}$$=$377.31/368~DOF is obtained at the grid point of
($\Delta m^2$, $\sin^2\theta_{23}$, $\sin^2\theta_{13}$)$=$
(-2.5$\times$10$^{-3}$eV$^2$, 0.525, 0.00625).
There is little difference in the $\chi^2$$_{min}$ values of the
normal and inverted hierarchy cases;
therefore both hypotheses are allowed by Super-K data.
Figure \ref{fig:sin2th13} shows $\chi^2-\chi^2$$_{min}$ distributions
projected to $\sin^2\theta_{13}$, 
in which minimum $\chi^2$ values for each $\Delta m^2$ and $\sin^2\theta_{23}$
are plotted.
It is shown that
the $\chi^2-\chi^2$$_{min}$ distribution for the inverted hierarchy 
is flatter and a larger $\sin^2\theta_{13}$ value is allowed.
The constraint on $\sin^2\theta_{13}$ is weaker for the inverted hierarchy case; 
$\sin^2\theta_{13}<0.27$ 
and 0.37$<$$\sin^2 \theta_{23}$$<$0.69
are allowed at 90\% confidence level.

The present analysis obtained upper limits on \(\theta_{13}\) which 
confirms CHOOZ experiment~\cite{Apollonio:1999ae}
(shown by Fig. \ref{fig:allowed} and \ref{fig:allowed_inv})
and Palo Verde~\cite{Piepke:2002ju}.
These limits are also consistent 
with the recent result by the K2K experiment~\cite{Yamamoto:2006ty}
giving the upper limit of 
\(\sin^2\theta_{13} \sim 0.06\) 
at \(\Delta m^2=2.8\times10^{-3} \mbox{eV}^2\), 
assuming \(\sin^2\theta_{23}=0.5\).

\section{\label{sec:conc}Conclusion}
In summary, a three-flavor oscillation analysis assuming one mass scale
dominance ($\Delta m_{12}^2$ $=$0) was performed with Super-Kamiokande~I
FC$+$PC$+$UP$\mu$ combined dataset.
A multi-ring $e$-like sample, selected using a likelihood
method, was newly introduced to increase
the statistics of electron neutrinos and improve
the sensitivity to $\theta_{13}$.
The best-fitted parameters for three-flavor oscillation becomes
($\Delta m^2$, $\sin^2\theta_{23}$, $\sin^2\theta_{13}$)$=$
(2.5$\times$10$^{-3}$~eV$^2$, 0.5, 0.0) and
the region of $\sin^2\theta_{13}$$<$0.14 and
0.37$<$$\sin^2 \theta_{23}$$<$0.65 is allowed
at 90~\% confidence level, assuming normal mass hierarchy.
We also tested the inverted mass hierarchy case:
a wider region, $\sin^2\theta_{13}$$<$0.27 and
0.37$<$$\sin^2 \theta_{23}$$<$0.69 is allowed
at 90~\% confidence level.
Both mass hierarchy hypotheses agree with our data.
We obtained the upper limit on \(\theta_{13}\),
which is consistent with CHOOZ, Palo Verde, and K2K experiments,
by using high statistics atmospheric neutrinos.
In contrast to these past experiments,
the earth matter effect plays important role in the analysis.

\par

We gratefully acknowledge the cooperation of the Kamioka Mining and
Smelting Company.  The \sk experiment has been built and
operated from funding by the Japanese Ministry of Education, Culture,
Sports, Science and Technology, the United States Department of Energy,
and the U.S. National Science Foundation.
Some of us have been supported by funds from the Korean Research
Foundation (BK21) and the Korea Science and Engineering Foundation, 
the Polish Committee for Scientific Research (grant 1P03B08227),
Japan Society for the Promotion of Science, and
Research Corporation's Cottrell College Science Award.

\bibliography{references}

\subsection{Appendix}
\label{binned:data}

Table~\ref{tb:eventnumber1} summarizes the number of observed and 
expected FC, PC and UP\(\mu\) events for each bin. 
The Monte Carlo prediction does not include neutrino oscillations.
These binned data are used in the oscillation analysis.

\begin{table*}
 \begin{center}
\begin{tabular}{|l|r|r|r|r|r|r|r|r|r|r|}
  \multicolumn{11}{c}{FC single-ring e-like}  \\
 \hline
 $P_{lep}$ &I &II &III &IV &V & VI&VII &IIX &IX &X\\
 \hline
 1 &114(79.3) &95(83.3) &74(81.4) &94(82.0) &88(84.0) &91(79.8) &79(79.5) &74(84.2) &91(81.5) &100(82.9) \\
 2 &96(75.6) &93(71.7) &96(73.2) &90(69.4) &89(68.4) &85(68.8) &85(69.5) &74(67.2) &83(71.1) &78(69.7) \\
 3 &76(64.2) &80(66.9) &80(65.8) &69(63.6) &72(64.6) &60(64.1) &69(62.4) &71(61.7) &85(59.7) &63(57.5) \\
 4 &48(45.4) &57(47.9) &62(50.1) &52(50.9) &60(51.6) &74(51.6) &55(50.8) &58(49.1) &60(46.5) &43(42.5) \\
 5 &26(21.7) &35(23.2) &31(25.1) &37(25.8) &24(25.6) &38(25.9) &34(25.0) &24(26.1) &21(23.6) &20(18.5) \\
 6 &33(29.3) &35(33.2) &41(34.9) &37(39.7) &46(42.8) &49(43.9) &49(40.7) &32(39.5) &36(32.0) &36(27.3) \\
 7 &10(13.8) &20(16.6) &15(18.4) &28(23.5) &36(26.6) &19(24.2) &28(22.0) &24(19.9) &18(17.2) &9(12.4) \\
 8 & 9(5.27)  &5(5.40)   &10(7.49)  &6(9.42)   &14(12.4) &11(11.8) &16(8.38)  &8(8.26)   &2(5.87)   &5(4.18) \\
 9 & 2(1.47)  &4(2.83)   &3(2.62)   &7(3.73)   &7(4.38)   &6(4.83)   &6(3.38)   &1(2.63)   &1(1.71)   &1(1.63) \\
 10& 2(0.86)  &2(0.86)   &0(1.63)   &1(1.26)   &6(2.40)   &1(2.16)   &4(1.68)   &1(1.79)   &1(1.23)   &3(1.26) \\
 \hline
 \multicolumn{11}{c}{FC single-ring $\mu$-like }  \\
 \hline
 $P_{lep}$ &I &II &III &IV &V & VI&VII &IIX &IX &X\\
 \hline
 1 & 36(54.7) &40(53.7) &39(54.4) &37(55.1) &35(55.8) &34(53.8) &35(53.5) &45(53.6) &48(52.6) &46(52.1) \\
 2 & 86(124) &77(123) &99(123) &86(122) &87(119) &80(120) &91(123) &85(118) &94(116) &76(121) \\
 3 & 94(119) &60(112) &81(113) &94(116) &87(113) &84(113) &116(113) &119(112) &97(108) &118(105) \\
 4 & 52(91.1) &48(88.0) &53(90.5) &53(91.0) &68(94.7) &68(91.1) &72(89.6) &81(88.2) &91(84.5) &86(82.9) \\
 5 & 27(43.4) &22(45.9) &22(44.9) &37(44.5) &25(47.0) &40(47.5) &41(47.9) &41(42.6) &46(44.0) &48(43.6) \\
 6 & 27(58.8) &35(57.3) &29(59.6) &32(61.6) &35(62.3) &57(63.2) &66(64.4) &69(59.6) &49(55.1) &56(54.0) \\
 7 & 4(26.1)  &10(24.9) &12(23.8) &15(27.3) &16(30.7) &15(28.3) &27(28.6) &16(24.6) &25(25.3) &33(26.8) \\
 8$\sim$10 & 3(4.61)   &1(4.21)   &1(3.11)   &2(3.66)   &3(3.36)   &1(3.14)   &2(3.17)   &2(4.19)   &4(4.16)   &4(4.04) \\
 \hline
 \multicolumn{11}{c}{FC multi-ring $e$-like }  \\
 \hline
 $E_{tot}$ &I &II &III &IV &V & VI&VII &IIX &IX &X\\
 \hline
 b &16(16.9)&18(18.1)&16(21.2)&22(26.9)&26(27.8)&31(27.6)&23(24.0)&19(22.1)&14(16.8)&11(16.0)\\
 c &9(13.4)&12(14.9)&13(18.9)&26(22.4)&21(28.2)&22(28.1)&15(22.2)&11(17.9)&10(14.2)&12(12.8)\\
 d &4(5.77)&3(6.40)&6(8.44)&7(11.3)&12(15.1)&14(16.0)&13(12.4)&3(7.28)&4(6.20)&6(4.77)\\
 e &2(2.17)&4(2.71)&1(2.83)&3(5.05)&5(7.61)&4(8.38)&6(5.19)&2(3.74)&3(1.87)&1(2.07)\\
 f &0(0.89)&1(1.11)&1(1.68)&1(1.75)&5(4.00)&5(4.70)&1(2.33)&1(0.98)&2(0.89)&2(1.21)\\
 \hline
 \multicolumn{11}{c}{FC multi-ring $\mu$-like }  \\
 \hline
 $E_{vis}$ &I &II &III &IV &V & VI&VII &IIX &IX &X\\
 \hline
 a &14(27.6) &8(31.2) &20(33.4) &14(33.7) &25(36.1) &16(35.6) &21(34.1) &32(32.9) &29(28.9) &29(29.1) \\
 b &11(33.2) &14(33.6)&16(36.7) &19(39.9) &20(43.9) &33(43.0) &28(40.7) &31(40.7) &30(33.8) &25(32.5) \\
 c &6(22.4)  &11(22.8)&11(23.5) &7(27.5)  &13(31.6) &20(28.8) &19(28.0) &17(24.8) &23(22.4) &19(20.4) \\
 d$\sim$f &1(7.68)   &4(6.88)  &4(6.73)   &2(8.28)   &8(10.8)  &16(10.9) &8(8.49)   &6(7.81)   &6(6.29)   &11(7.92) \\
 \hline

 \multicolumn{11}{c}{}  \\
 \multicolumn{11}{c}{OD stopping PC }  \\
 \hline
 $E_{vis}$ &I &II &III &IV &V & VI&VII &IIX &IX &X\\
 \hline
 a &5(4.26) &2(3.53) &2(3.67) &2(4.29) &1(4.98) &5(4.14) &5(4.26) &6(4.14) &9(3.72) &5(3.78) \\
 b &2(4.38) &2(7.23) &3(6.67) &4(5.18) &6(5.81) &2(6.01) &3(6.85) &5(6.76) &9(6.42) &5(4.29) \\
 c &4(4.18) &7(5.61) &3(6.95) &1(6.88) &2(6.02) &5(6.28) &7(5.89) &4(6.57) &4(6.55) &8(4.63) \\
 d$\sim$f &1(4.30)   &0(4.63)  &3(3.81)   &6(4.75)   &5(4.08)  &8(5.16) &4(4.62)   &6(4.33)   &8(3.56)   &2(2.90) \\
 \hline
 \multicolumn{11}{c}{OD through-going PC }  \\
 \hline
 $E_{vis}$ &I &II &III &IV &V & VI&VII &IIX &IX &X\\
 \hline
 a &5(9.90) &9(5.69) &10(8.48) & 9(11.2) &9 (13.5) &9(15.1) &11(13.3) &10(8.53) &10(7.13) &7(9.55) \\
 b &4(13.9) &6(14.0) &10(17.3) &21(22.1) &18(25.2) &18(25.9) &12(19.9) &22(17.9) &11(14.0) &20(14.7) \\
 c &8(18.7) &6(20.9) &12(26.1) &15(32.4) &20(36.9) &48(40.2) &36(34.5) &27(24.1) &11(21.8) &18(22.2) \\
 d$\sim$f &20(29.4)   &13(27.2)  &16(31.5)   &31(38.4)   &56(58.2)  &61(55.6) &36(39.1)   &29(30.7)   &23(26.5)   &23(26.2) \\
 \hline
 \multicolumn{11}{c}{}  \\
 \multicolumn{11}{c}{Upward stopping muon }  \\
\hline
    & 28(51.2) & 23(54.1) & 37(56.7) & 30(65.0) & 27(67.6) & 37(68.2) & 37(78.9) & 48(81.0) & 65(94.0) & 85.7(96.9) \\
 \hline
 \multicolumn{11}{c}{Upward through-going muon}  \\
 \hline
    & 85(96.1) & 113(115)  & 116(122) & 138(137) & 159(146) & 183(169) & 178(187) & 267(211) & 286(229) & 316.6(257) \\
 \hline
  \end{tabular} 
  \end{center} 
  \caption{Summary of the number of observed (MC expected) FC, PC and 
    UP\(\mu\) events  
    for each bin.  Neutrino oscillation is not included 
    in the MC prediction.  Roman numbers I, II, ... X represent zenith angle 
    regions  $-1 < \cos\Theta < -0.8 $, $-0.8 < \cos\Theta < -0.6 $, ... and
    $0.8 < \cos\Theta < 1.0$ respectively for FC and PC events, 
    $-1 < \cos\Theta < -0.9 $, $-0.9 < \cos\Theta < -0.8 $, ... and
    $0.1 < \cos\Theta < 0.0$ respectively for upward stopping and through-going muon events. 
    The numbers 1 to 5 in the $P_{lep}$ column correspond to the momentum ranges 
     $<$250, 250-400, 400-630, 630-1000 and $>$1000~MeV/c for sub-GeV samples 
   and the numbers 6 to 10 correspond to
     $<$2.5,  2.5-5.0, 5.0-10, 10-20~ and $>$20~GeV/c for multi-GeV samples.
    The letters a to f in the $E_{tot}$ and $E_{vis}$ columns correspond to energy ranges 
    0.2-1.33, 1.33-2.5, 2.5-5.0, 5.0-10, 10-20, $>$20~GeV.
    }
\label{tb:eventnumber1} 
\end{table*}


\end{document}